\documentclass[aps,prd,reprint,twocolumn,nofootinbib,superscriptaddress]{revtex4-1}
\pdfoutput=1 
\usepackage{graphicx}% Include figure files
\usepackage{dcolumn}% Align table columns on decimal point
\usepackage{bm}% bold math
\usepackage{color}
\usepackage{textcomp}
\usepackage{subfigure}
\usepackage{feynmp}
\usepackage{slashed}
\usepackage{amsmath,amssymb,array}
\usepackage{enumerate}
\usepackage{hhline}
\usepackage{booktabs}
\usepackage{multirow}
\usepackage{siunitx}
\usepackage{orcidlink}
\newcommand{\cc}{C}
\newcommand{\beqa}{\begin{eqnarray}}
\newcommand{\eeqa}{\end{eqnarray}}
\newcommand{\be}{\begin{equation}}
\newcommand{\ee}{\end{equation}}
\newcommand{\ba}{\begin{array}} 
\newcommand{\ea}{\end{array}}

\begin{document} 
\title{Quantum corrections and the minimal Yukawa sector of $SU(5)$}
\author{Ketan M. Patel\orcidlink{0000-0002-6889-7470}}
\email{ketan.hep@gmail.com}
\affiliation{Theoretical Physics Division, Physical Research Laboratory, Navarangpura, Ahmedabad-380009, India}
\author{Saurabh K. Shukla\orcidlink{0000-0001-5344-9889}}
\email{saurabhks@prl.res.in}
\affiliation{Theoretical Physics Division, Physical Research Laboratory, Navarangpura, Ahmedabad-380009, India}
\affiliation{Indian Institute of Technology Gandhinagar, Palaj-382055, India}

\begin{abstract}
It is well-known that the $SU(5)$ grand unified theory, with the standard model quarks and leptons unified in $\overline{5}$ and $10$ and the electroweak Higgs doublet residing in $5$ dimensional representations, leads to relation, $Y_d=Y_e^T$, between the Yukawa couplings of the down-type quarks and the charged leptons. We show that this degeneracy can be lifted in a phenomenologically viable way when quantum corrections to the tree-level matching conditions are taken into account in the presence of one or more copies of gauge singlet fermions. The 1-loop threshold corrections arising from heavy leptoquark scalar and vector bosons, already present in the minimal model, and heavy singlet fermions can lead to realistic Yukawa couplings provided their masses differ by at least two orders of magnitude. The latter can also lead to a realistic light neutrino mass spectrum through the type I seesaw mechanism if the colour partner of the Higgs stays close to the Planck scale. Most importantly, our findings demonstrate the viability of the simplest Yukawa sector when quantum corrections are considered and sizeable threshold effects are present.
\end{abstract}

\maketitle

\section{Introduction}
\label{sec:intro}
After the remarkable realization of the potential unification of the standard model (SM) gauge symmetries into a single gauge symmetry nearly fifty years ago \cite{Georgi:1974sy,Fritzsch:1974nn,Pati:1974yy}, it has since become well-established that the Yukawa sector of the SM plays a pivotal role in determining the minimal and viable configurations of grand unified theories (GUT). The latter's potential to partially or completely unite quarks and leptons, in conjunction with the simplest choice of the Higgs field(s) in the Yukawa sector, often results in correlations among the effective SM Yukawa couplings that are inconsistent with observations.

The most glaring and simplest example of the above is the $SU(5)$ GUTs with only $5$-dimensional ($5$ and $\overline{5}$) Lorentz scalar(s) in the Yukawa sector in their ordinary (supersymmetric) versions. Both lead to
\be \label{Yd_Ye_tree}
Y_{d} = Y_e^T\,,\ee
at the scale of the unified symmetry breaking, namely $M_{\rm GUT}$, for the down-type quark and charged-lepton Yukawa coupling matrices $Y_d$ and $Y_e$, respectively. The degeneracy between the two sectors predicted by Eq. (\ref{Yd_Ye_tree}) is not supported by the GUT scale extrapolated values of the effective Yukawa couplings determined from the measured masses of the down-type quarks and the charged leptons \cite{Buras:1977yy,Langacker:1980js}. The largest mismatch arises in the case of non-supersymmetric theories in which the extrapolation of the SM data implies, $y_b/y_\tau \approx 2/3$, $y_s/y_\mu \approx 1/5$ and $y_d/y_e \approx 2$, at $M_{\rm GUT} = 10^{16}$ GeV.

Deviation from the degeneracy shown in Eq. (\ref{Yd_Ye_tree}) can be achieved through several means: (a) Expanding the scalar sector \cite{Georgi:1979df,Barbieri:1981yw,Giveon:1991zm,Dorsner:2006dj,FileviezPerez:2007bcw,Goto:2023qch}, for instance, by introducing a $45$-dimensional Higgs field, or (b) Incorporating higher-dimensional non-renormalizable operators \cite{Ellis:1979fg,Berezinsky:1983va,Altarelli:2000fu,Emmanuel-Costa:2003szk,Dorsner:2006hw,Antusch:2021yqe,Antusch:2022afk}, or (c) Introducing vector-like fermions that mix with the charged leptons and/or down-type quarks residing in the chiral multiplets of $SU(5)$ \cite{Hempfling:1993kv,Shafi:1999rm,Barr:2003zx,Oshimo:2009ia,Babu:2012pb,Dorsner:2014wva,FileviezPerez:2018dyf,Antusch:2023mxx,Antusch:2023kli,Antusch:2023mqe}. Each of these approaches alters the tree-level matching condition, Eq. (\ref{Yd_Ye_tree}), and introduces new couplings. These new couplings can be harnessed to obtain effective Yukawa couplings compatible with the SM.

In this article, we present a rather simple approach to alleviate the degeneracy between charged leptons and down-type quarks. Our method involves incorporating higher-order corrections to the tree-level matching conditions for the Yukawa couplings. Non-trivial implications of such corrections  in the context of supersymmetric versions of $SO(10)$ GUTs have been pointed out in \cite{Aulakh:2006hs,Aulakh:2008sn,Aulakh:2013lxa}\footnote{Nevertheless, the degeneracy, as in Eq. (\ref{Yd_Ye_tree}), is absent in these models even at the tree level due to the presence of multiple scalars containing the SM Higgs doublets.}.  In the context of $SU(5)$, we show that the inclusion of such corrections does not necessitate the introduction of new fermions or scalars charged under the $SU(5)$ for modifying the tree-level Yukawa relations. This sets the present proposal apart from the previous ones outlined as (a-c) above.  Specifically, we demonstrate that by expanding the minimal non-supersymmetric $SU(5)$ framework to include fermion singlets and accounting for threshold corrections to the Yukawa couplings originating from these singlets, along with the leptoquark scalar and vector components already present in the minimal setup,  a fully realistic fermion spectrum can be achieved.

\section{Yukawa relations at 1-loop}
\label{sec:Yuk}
The Yukawa sector of the model is comprised of three generations of ${\bf 10}$, $\overline{\bf 5}$ and $N$ generations of the gauge singlet ${\bf 1}$ Weyl fermions and a Lorentz scalar $5_H$. The most general renormalizable interactions between these fields can be parametrized as
\beqa \label{LY}
{\cal L}_{\rm Y} &=& \frac{1}{4} (Y_1)_{ij} {\bf 10}_{i}^T \cc {\bf 10}_j 5_H + \sqrt{2} (Y_2)_{ij} {\bf 10}_{i}^T\cc \overline{\bf 5}_j 5^*_H \nonumber \\
& + & (Y_3)_{i\alpha} {\bf{\overline{5}}}_i^T \cc {\bf 1}_\alpha 5_H  -\frac{1}{2} (M_N)_{\alpha \beta}{\bf 1}^T_\alpha C {\bf 1}_\beta + {\rm h.c.}\,, \eeqa
with $i,j=1,2,3$ and $\alpha=1,...,N$ denotes the generations and $C$ is the usual charge-conjugation matrix. We have suppressed the gauge and Lorentz indices for brevity. The symmetric nature of the first term implies $Y_1 = Y_1^T$. Additionally, $M_N$ is the gauge invariant Majorana mass of the singlet fermions alias the right-handed (RH) neutrinos.
%\beqa \label{MR}
%{\cal L}_{M} = -\frac{1}{2} (M_N)_{\alpha \beta} {\bf 1}^T_\alpha C {\bf 1}_\beta + {\rm %h.c.}\,.\eeqa 

The SM quarks and leptons residing in the $SU(5)$ multiplets are identified as ${\bf 10}^{ab} = \frac{1}{\sqrt{2}} \epsilon^{abc} u^C_c$, ${\bf 10}^{am} = -\frac{1}{\sqrt{2}} q^{am}$,  ${\bf 10}^{mn} = -\frac{1}{\sqrt{2}} \epsilon^{mn} e^C$, $\overline{\bf 5}_a = d^C_a$, $\overline{\bf 5}_m = \epsilon_{mn} l^n$ and ${\bf 1} = \nu^C$, where $a,b,c$ denote the color while $m,n$ are $SU(2)$ indices.  For the scalar, we define a colour triplet $T^a \equiv 5_H^a$ and an electroweak doublet $h^m \equiv 5_H^m$ \cite{Patel:2022wya}. Decompositions of Eq. (\ref{LY}) then lead to the following Yukawa interactions with the colour triplet and Higgs:
\beqa \label{L_T}
-{\cal L}_{\rm Y}^{(T)} &=& (Y_1)_{ij} \left(u^{CT}_i C e^C_j + \frac{1}{2} q^T_i C q_j\right) T \nonumber \\
& - & (Y_3)_{i\alpha}\, d^{CT}_i C \nu^C_\alpha T \nonumber \\
&-& (Y_2)_{ij}  \left(u^{CT}_i C d^C_j + q^T_i C l_j\right) T^* + {\rm h.c.}\,,\eeqa
and 
\beqa \label{L_h}
-{\cal L}_{\rm Y}^{(h)} &=& (Y_1)_{ij} q^T_i C u^C_j \tilde{h} + (Y_2)_{ij} q^T_i C d^C_j h^* \nonumber \\
&+& (Y_3)_{i\alpha} l^T_i C \nu^C_\alpha \tilde{h} + (Y_2^T)_{ij} l^T_i C e^C_j h^* + {\rm h.c.}\,,\eeqa
where $\tilde{h} = \epsilon h$ and we have suppressed the $SU(3)$ and $SU(2)$ contractions. Matching of ${\cal L}_{\rm Y}^{(h)}$ with the SM Yukawa Lagrangian at tree level leads to $Y_u=Y_1$ and $Y_d = Y_e^T = Y_2$  at the renormalization scale  $\mu=M_{\rm GUT}$.

For the matching at 1-loop, the Yukawa couplings receive two types of contributions. The first arises from the vertex corrections involving the colour triplet or the leptoquark gauge boson in the loop. The interaction of the latter with the SM fermions originates from the unified gauge interaction and it is given by \cite{Buras:1977yy,Langacker:1980js}
\beqa \label{L_X}
-{\cal L}_{\rm G}^{(X)} &=& \frac{g}{\sqrt{2}} \overline{X}_\mu \left( \overline{d^C}_i \overline{\sigma}^\mu l_i - \overline{q}_i \overline{\sigma}^\mu u^C_i - \overline{e^C}_i \overline{\sigma}^\mu q_i \right) + {\rm h.c.}\,, \nonumber \\
\eeqa
where $X$ transforms as $(3,2,-5/6)$ under the SM gauge symmetry. The second type of contribution to the Yukawa threshold correction is due to wavefunction renormalization of fermions and scalar involving at least one of the heavy fields in the loop.

The 1-loop corrected matching condition for the Yukawa couplings at a renormalization scale $\mu$ is given by 
\beqa \label{dY_gen}
Y_f = Y_f^{0}\left(1-\frac{K_h}{2}\right) + \delta Y_f - \frac{1}{2} \left(K^T_f Y_f^{0} + Y_f^{0} K_{f^C}\right),\eeqa
where $f=u,d,e,\nu$. The details of the derivation of the above expression are outlined in Appendix \ref{app:gen_form}. In Eq (\ref{dY_gen}), $\delta Y_f$ are the finite parts of 1-loop corrections to the Yukawa vertex $Y_f$ while $K_{f,f^C,h}$ are the finite parts of the wavefunction renormalization diagrams involving heavy particles in the loops evaluated in the $\overline{\rm MS}$ scheme. $Y_f^0$ denotes the tree-level Yukawa coupling matrix. As mentioned earlier,
\be \label{Y0}
Y_u^0 = Y_1\,,~~Y_d^0 = Y_2\,,~~Y_e^0 = Y_2^T\,,~~Y_\nu^0 = Y_3\,, \ee 
at $\mu = M_{\rm GUT}$.

Next, we compute $\delta Y_f$ using the interaction terms given in Eqs. (\ref{L_T},\ref{L_h},\ref{L_X}) and assuming massive color triplet scalar $T$, vector leptoquark $X$ and $N$ generations of the RH neutrinos $\nu^C_\alpha$. We find,
\beqa \label{dY}
(\delta Y_u)_{ij} &=& 4 g^2 (Y_1)_{ij} f[M_X^2,0] \nonumber \\ 
&+& \left(Y_1 Y_2^* Y_2^T + Y_2 Y_2^\dagger Y_1^T \right)_{ij} f[M_T^2,0], \nonumber\\ 
(\delta Y_d)_{ij} &=& 2 g^2 (Y_2)_{ij} f[M_X^2,0]+\left(Y_1 Y_1^* Y_2 \right)_{ij} f[M_T^2,0]\nonumber \\
& + & \sum_\alpha \left(Y_2 Y_3^* \right)_{i \alpha} \left(Y_3^T\right)_{\alpha j} f[M_T^2,M_{N_\alpha}^2], \nonumber\\
(\delta Y_e)_{ij} &=& 6 g^2 (Y_2^T)_{ij} f[M_X^2,0]+ 3 \left(Y_2^T Y_1^* Y_1 \right)_{ij} f[M_T^2,0], \nonumber\\
(\delta Y_\nu)_{i\alpha} &=& 3 \left(Y_2^T Y_2^* Y_3 \right)_{i\alpha} f[M_T^2,0],\eeqa
at the scale $\mu$. Here, $M_{N_\alpha}$ is the mass of $\nu^C_\alpha$ and $f[m_1^2,m_2^2]$ is a loop integration factor and it is given in Eq. (\ref{LF_f}) in the Appendix \ref{app:LF}. It can be noticed that other than the overall colour factor, $\delta Y_d$ and $\delta Y_e$ differ by the contribution from the heavy RH neutrinos. Because of the tree-level Yukawa couplings between $d^C_i$, $\nu^C_\alpha$ and $T$ in Eq. (\ref{L_T}), the $Y_d$ gets threshold correction from the RH neutrinos and colour triplet scalar. It is noteworthy that the corrections $\delta Y_f$ vanish in the supersymmetric version of the model \cite{Wright:1994qb}, due to the perturbative non-renormalisation theorem for the supersymmetric field theories \cite{Grisaru:1979wc,Seiberg:1993vc}.

The computations of the finite parts of wavefunction renormalization for the light fermions and scalar at 1-loop, involving at least one heavy fields in the loop, lead to:
\beqa \label{K}
(K_{q})_{ij} &=& 3g^2 \delta_{ij} h[M_X^2,0] \nonumber \\
&-& \frac{1}{2}\left( Y_1^{*} Y_1^T +2  Y_2^* Y_2^{T}\right)_{ij} h[M_T^2,0], \nonumber\\
(K_{u^C})_{ij} &=& 4g^2 \delta_{ij} h[M_X^2,0] \nonumber \\
&-& \left(Y_1^* Y_1^T +2 Y_2^* Y_2^T\right)_{ij} h[M_T^2,0], \nonumber\\
(K_{d^C})_{ij} &=& 2g^2 \delta_{ij} h[M_X^2,0] - 2 \left(Y_2^\dagger Y_2\right)_{ij} h[M_T^2,0] \nonumber \\
& - & \sum_\alpha \left(Y_3^*\right)_{i \alpha} \left(Y_3^T\right)_{\alpha j} h[M_T^2,M_{N_\alpha}^2], \nonumber\\
(K_l)_{ij} &=& 3g^2 \delta_{ij} h[M_X^2,0] - 3 \left(Y_2^\dagger Y_2\right)_{ij} h[M_T^2,0], \nonumber\\
(K_{e^C})_{ij} &=& 6g^2 \delta_{ij} h[M_X^2,0] - 3 \left(Y_1^\dagger Y_1\right)_{ij} h[M_T^2,0], \nonumber\\
(K_{\nu^C})_{\alpha \beta} &=& - 3 \left(Y_3^\dagger Y_3\right)_{\alpha \beta} h[M_T^2,0], \nonumber\\
K_{h} &=& \frac{g^2}{2}\,\left(f[M_X^2,M_T^2] + g[M_X^2,M_T^2]\right), \eeqa 
at the scale $\mu$. The loop integration factors are defined in Appendix \ref{app:LF}. Again, only $K_{d^C}$ receives a contribution from the singlet fermions. As we show in the next sections, these contributions from singlet fermions are crucial for uplifting degeneracy between the charged lepton and down-type quarks.

\section{Deviation from $Y_d = Y_e^T$}
\label{dep_ydye}
It is seen from Eqs. (\ref{dY_gen},\ref{dY},\ref{K}) that the 1-loop corrections break the degeneracy between $Y_e$ and $Y_d$. Explicitly, we obtain at the GUT scale:
\beqa  \label{ydye_ana}
\left(Y_d - Y_e^T\right)_{ij} &=& -2g^2 (Y_2)_{ij}\, \left(2f[M_X^2,0] - h[M_X^2,0] \right) \nonumber \\
&-& \left(Y_1 Y_1^* Y_2\right)_{ij}\, \left(f[M_T^2,0] +\frac{5}{8} h[M_T^2,0] \right) \nonumber \\
&+& \sum_\alpha \left(Y_2 Y_3^*\right)_{i \alpha} \left(Y_3\right)_{j \alpha} \Big(f[M_T^2,M_{N_\alpha}^2] \Big. \nonumber \\
&+& \Big. \frac{1}{2} h[M_T^2,M_{N_\alpha}^2] \Big) \,.\eeqa
The above is the main result of this paper. It is noteworthy that Eq. (\ref{ydye_ana}) not only suggests $Y_d \neq Y_e^T$ but also implies that the difference between the two matrices is calculable in terms of the masses of the heavy scalar, gauge boson and RH neutrinos and their couplings. The latter also determines the masses of other fermions and hence can be severely constrained as we discuss in the next section.

\begin{figure}[t]
\centering
\subfigure{\includegraphics[width=0.40\textwidth]{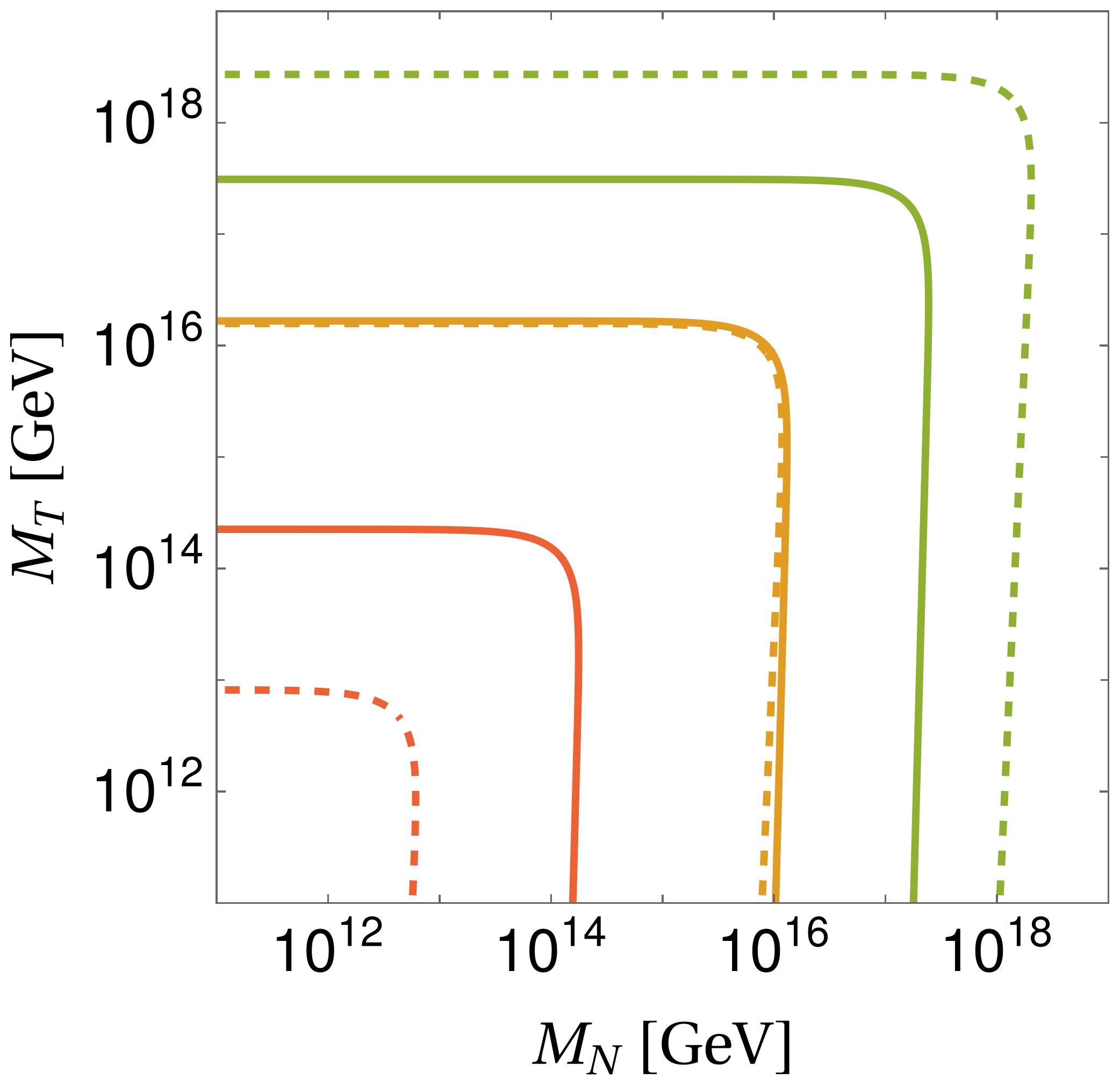}}
\caption{Contours of $y_b/y_\tau = 3/2$ (red), $y_b/y_\tau = 1$ (orange) and $y_b/y_\tau = 2/3$ (green) drawn using Eq. (\ref{btau_ana}) for $y_t=0.427$, $g=0.53$ and $\mu=M_X=10^{16}$ GeV and for $y_\nu = \sqrt{4 \pi}$ (solid lines) and $y_\nu = 2.7$ (dashed lines).}
\label{fig1}
\end{figure}

Before assessing the viability of Eq. (\ref{ydye_ana}) in reproducing the complete and realistic fermion mass spectrum, we investigate its role for the third generation Yukawa couplings, namely $y_b$ and $y_\tau$, through a simplified analysis. Considering only one RH neutrino with $M_{N_1} = M_N$ and only the third generation, one finds from Eq. (\ref{ydye_ana}):
\beqa  \label{btau_ana}
\frac{y_b}{y_\tau} &\simeq& 1-2g^2 \left(2f[M_X^2,0] - h[M_X^2,0] \right) \nonumber \\
&-& 2y_t^2 \left(f[M_T^2,0] +\frac{5}{8} h[M_T^2,0] \right) \nonumber \\
&+& y_\nu^2 \left(f[M_T^2,M_N^2] +\frac{1}{2} h[M_T^2,M_N^2] \right)\,,\eeqa
at the GUT scale. Here, $y_t$ is the top-quark Yukawa coupling and $y_\nu = (Y_3)_{31}$. For some sample values of $y_t$, $y_\nu$ and $\mu=M_X=10^{16}$ GeV, the contours corresponding to different values of the ratio $y_b/y_\tau$ on the $M_T$-$M_N$ plane are displayed in Fig. \ref{fig1}.

%%%%%%%%%%%%%%%
%%%%%%%%%%%%%%
The GUT scale extrapolation of the observed fermion mass data requires $y_b/y_\tau \approx 2/3$. As can be seen from Fig. \ref{fig1}, this can be achieved only if either $M_T$ or $M_N$ is larger than $\mu=M_X$ by at least one to two orders of magnitude. Moreover, $y_\nu$ is also required to be large. For $g,y_t < 1$, it is the third term in Eq. (\ref{btau_ana}) which is required to dominantly contribute to uplift the degeneracy between $y_b$ and $y_\tau$ and hence the largest possible value of $y_\nu$ is preferred. $M_T \gg M_{\rm GUT}$ or $M_N \gg M_{\rm GUT}$ along with large $y_\nu$ are needed to overcome the loop suppression factor of $1/(16 \pi)^2$. This simple picture provides a clear and qualitative understanding of the favourable mass scales of the colour triplet scalar and RH neutrino and it also holds more or less when the full three generation fermion spectrum is considered as we show in the next section.

It is noteworthy that the RH neutrino through its coupling with the lepton doublet generates a contribution to the light neutrino mass through the usual type I seesaw mechanism \cite{Minkowski:1977sc,Yanagida:1979as,Mohapatra:1979ia,Schechter:1980gr}. It is obtained as $m_\nu = v^2 y_\nu^2/M_N$. If this contribution is required to generate the atmospheric neutrino mass scale then one finds,
\be \label{MN_1gen}
M_N = 7.6 \times 10^{16}\,{\rm GeV}\,\left(\frac{y_\nu}{\sqrt{4 \pi}}\right)^2\,\left(\frac{0.05\,{\rm eV}}{m_\nu} \right)\,. \ee
Since $M_N$ cannot be much larger than $M_{\rm GUT}$ in this case, phenomenologically viable $y_b/y_\tau$ can be achieved only if $M_T > M_{{\rm GUT}}$. Conversely, when considering perturbative values of $y_\nu$ and a situation where $M_N$ greatly surpasses $M_{\rm GUT}$, the RH neutrino's contribution to the light neutrino mass is rather negligible. This inadequacy to reproduce a viable atmospheric neutrino mass scale necessitates the inclusion of an additional source of neutrino masses. We also provide an example of this in the next section.

\section{Viability test and results}
\label{sec:res}
To establish if the $Y_u$, $Y_d$ and $Y_e$ are evaluated from Eqs. (\ref{dY_gen},\ref{dY},\ref{K}) can reproduce the realistic values of the SM Yukawa couplings and the quark mixing (CKM) matrix, we carry out the $\chi^2$ optimization. Focusing on the minimal  setup, we first consider only one RH neutrino with mass $M_{N_1} \equiv M_N$ as mentioned in the previous section. The $\chi^2$ function (see for example \cite{Joshipura:2011nn,Mummidi:2021anm} for the definition and optimization procedure) includes 9 diagonal charged fermion Yukawa couplings and 4 CKM parameters. For the input values of these parameters at the GUT scale, we evolve the SM Yukawa couplings from $\mu=M_t$ ($M_t$ being the top pole mass)  to $\mu= M_{\rm GUT} = 10^{16}$ GeV using the 2-loop renormalization group equations (RGEs) in the $\overline{\rm MS}$ scheme following the procedure outlined in \cite{Mummidi:2021anm}. The 2-loop SM RG equations have been computed using the {\sc PyR$@$TE} 3 package \cite{Sartore:2020gou}. The values of the SM Yukawa and gauge couplings at $\mu=M_t$ are taken from \cite{Surya:2020ydm}. The RGE extrapolated values at the GUT scale are listed as $O_{\rm exp}$ in Table \ref{tab:tab1}. For the standard deviations, we use $\pm 30 \%$ in the light quark Yukawa couplings $(y_{u,d,s})$ and $\pm 10\%$ in the rest of the observables as considered in the previous fits \cite{Mummidi:2021anm}.

\begin{figure}[t]
\centering
\subfigure{\includegraphics[width=0.40\textwidth]{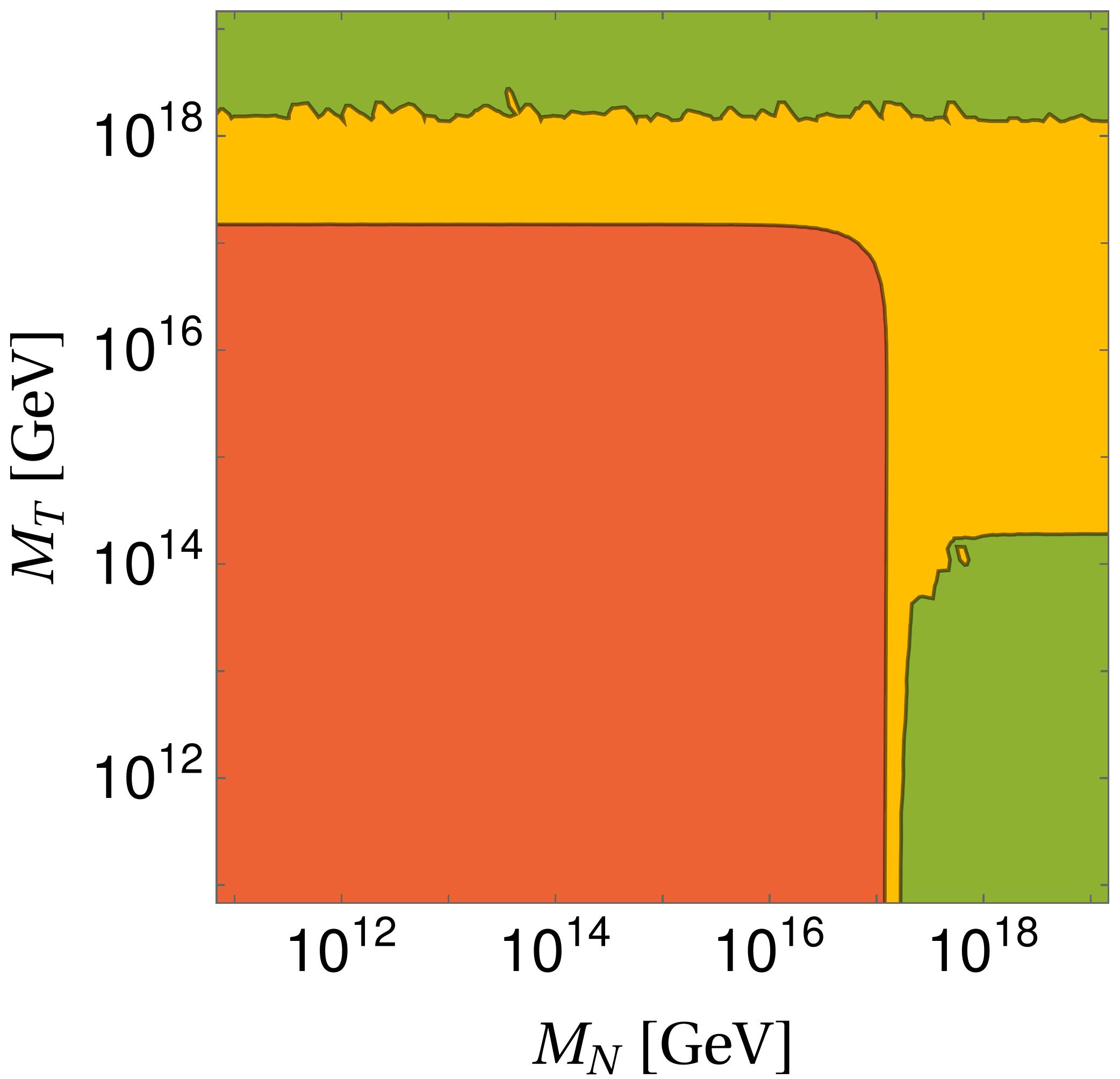}}
\caption{The distribution of minimized $\chi^2$ for different values of $M_T$ and $M_N$. The green, yellow and red regions correspond to $\chi^2_{\rm min} \leq 3$, $3<\chi^2_{\rm min}  \leq 9$ and $\chi^2_{\rm min} > 9$, respectively. For the fits, we set $\mu=M_X=10^{16}$ GeV,  $g=0.53$ and impose $|(Y_{1,2,3})_{ij}| < \sqrt{4 \pi}$.} 
\label{fig2}
\end{figure}

Using the freedom to choose a basis in Eq. (\ref{LY}), we set $Y_1$ diagonal and real. The RH neutrino mass matrix, $M_N$, in general $N$ flavour case can also be chosen real and diagonal simultaneously. $Y_{2,3}$ are complex in this basis. Using the Eqs. (\ref{dY_gen},\ref{dY},\ref{K}), we then compute the matrices $Y_{u,d,e}$ and diagonalize them to obtain the nine diagonal Yukawa couplings and quark mixing parameters. These quantities are fitted to the extrapolated data at $\mu=M_{\rm GUT}$ by minimizing the $\chi^2$ function. We set $M_X=M_{\rm GUT}$ and $g=0.53$ which is an approximate value of the RGE evolved SM gauge couplings at $\mu=10^{16}$ GeV. Fixing $M_T$ and $M_N$ to some values, we then minimize the $\chi^2$ along with a constrain $|(Y_{1,2,3})_{ij}| < \sqrt{4 \pi}$ on all the input Yukawa couplings to ensure that they are within the perturbative limits \cite{Allwicher:2021rtd}. We repeat this procedure for several values of $M_T$ and $M_N$. The obtained distribution of the minimized $\chi^2$ ($\equiv \chi^2_{\rm min}$) is displayed in Fig. \ref{fig2}. 
%%%%%%%%%%%%%%%
%%%%%%%%%%%%%%

Note that without 1-loop corrections, i.e. with $Y_d = Y_e^T$, the obtained value of $\chi^2_{\rm min}$ is 53. Therefore, values of $\chi^2_{\rm min} < 53$ show improvements due to quantum corrected matching conditions in the model. In particular, for $\chi^2_{\rm min} < 9$, it is ensured that no observable is more than $\pm 3\sigma$ away from its central value and, therefore, can be considered to lead to viable charged fermion mass spectrum and the quark mixing.

It can be seen from Fig. \ref{fig2}, a very good fit of the entire charged fermion mass spectrum and the quark mixing parameters can be obtained if $M_T $ or $M_N \geq 10^{17.2}$ GeV. These results are in a very good  agreement with the limits on $M_T$ and $M_N$ obtained for $y_b/y_\tau \lesssim 2/3$ in a simplified case discussed earlier and shown in Fig. \ref{fig1}.

The three-generation $\chi^2$ analysis also reveals that all the underlying 13 observables can be fitted within their $\pm 1\sigma$ range (corresponding to $\chi_{\rm min}^2 \leq 3$) provided (i) $M_T \leq 10^{14.5}$ GeV and $M_N \geq 10^{17.2}$ GeV, or (ii) $M_T \geq 10^{18.2}$ GeV. While the second leads to $M_T$ alarmingly close to the Planck scale making the doublet-triplet splitting problem \cite{Masiero:1982fe,Grinstein:1982um,Babu:2006nf} more severe, the possibility (i) is conceptually allowed and technically a safe choice.  Since $M_N$ is a scale independent of $M_{\rm GUT}$ in the present framework, the large hierarchy between them is permitted. Also, $M_T$ can be significantly smaller than $M_{\rm GUT}$ provided it satisfies the proton lifetime limit, $M_T \gtrsim 10^{11}$ GeV \cite{Dorsner:2012uz}. We list explicitly one benchmark solution from the region (i) which is displayed as Solution I in Table \ref{tab:tab1}. The fitted values of the corresponding input parameters are given in Appendix \ref{app:sol}.

%%%%%%%%%%%%%%%%%%%%
\begin{table*}[t]
	\begin{center} 
		\begin{math} 
			\begin{tabular}{cccccccc}
				\hline
				\hline 
      \multirow{2}{*}{~~~~~Observable~~~~~} & \multirow{2}{*}{ \hspace*{0.75cm}  $O_{\rm exp}$ \hspace*{0.75cm} } &
      \multicolumn{2}{c}{\bf \hspace*{0.75cm} Solution I \hspace*{0.75cm} } & 
      \multicolumn{2}{c}{\bf \hspace*{0.75cm}  Solution II \hspace*{0.75cm} } &
      \multicolumn{2}{c}{\bf \hspace*{0.75cm}  Solution III \hspace*{0.75cm} }\\
				 &   & $O_{\rm th}$ & pull &$O_{\rm th}$ & pull & $O_{\rm th}$ & pull \\
				\hline
				$y_u$  & $2.81 \times 10^{-6}$ & $2.92\times 10^{-6}$ & $ 0$ &$2.81 \times10^{-6}$& $ 0$ & $2.81\times 10^{-6}$ & $0$ \\
				$y_c$   & $1.42\times10^{-3}$& $1.42\times 10^{-3}$& $0$&$1.42\times10^{-3}$& $0$ & $1.42 \times 10^{-3}$ & $0$\\
				$y_t$   & $4.27 \times 10^{-1} $ & $4.35 \times 10^{-1}$& $0.2$& $4.30\times10^{-1}$& $0.1$ & $4.28 \times 10^{-1}$& $\sim 0$\\
				$y_d$  & $ 6.14 \times 10^{-6} $ & $3.60\times10^{-6}$  & $-1.2$& $ 2.85\times 10^{-6}$ & $-1.8$ & $2.91 \times 10^{-6}$ & $-1.8$\\
				$y_s$   & $1.25 \times10^{-4}$ & $1.26\times10^{-4}$ & $\sim 0$ & $1.24\times 10^{-4}$ & $\sim 0$ & $1.25\times 10 ^{-4}$ & $\sim 0$\\
				$y_b$   & $5.80 \times10^{-3}$ & $5.77 \times10^{-3}$& $\sim 0$& $6.09\times 10^{-3}$& $0.5$ & $5.79\times 10^{-3}  $ & $\sim 0$\\
				$y_e$   & $2.75 \times 10^{-6}$ & $2.76 \times 10^{-6}$ & $0.2$ & $2.81 \times 10^{-6}$& $0.2$ & $2.82 \times 10^{-6}$ & $0.3$\\
				$y_{\mu}$  &$5.72 \times10^{-4}$ & $5.71\times10^{-4}$ & $\sim0$& $5.65 \times 10^{-4}$ & $-0.1$ & $5.71 \times 10^{-4}$& $\sim 0 $\\
				$y_{\tau}$   & $9.68 \times10^{-3}$ & $9.83 \times10^{-3}$& $ 0.2$& $9.06 \times 10^{-3}$ & $-0.6$ & $9.70\times 10^{-3}$ & $\sim 0$\\
				$|V_{us}|$ & $0.2286$ & $0.2303$ & $0.1$ & $0.2292$& $\sim 0$ & $0.2291$ & $\sim 0$\\
				$|V_{cb}|$ & $0.0457$ & $0.0461$ & $0.1$& $0.0458$& $\sim 0$ & $0.0458$ & $\sim 0$\\
				$|V_{ub}|$ & $0.004$2 & $0.0043$ & $\sim 0$ & $0.0042$ & $\sim 0$ & $0.0042$ & $\sim 0$ \\
				$\sin\delta_{\rm CKM}$ & $0.78$ & $0.78$ & $ 0$ & $0.78$ & $ 0$ & $0.78$ & $\sim 0$  \\
				$\Delta m^2_{\text{sol}}~ [{\rm eV}^2]$ & $7.41\times10^{-5}$& $-$ & $- $& $7.53 \times 10^{-5}$ & $\sim 0$ & $7.51\times 10^{-5}$ & $\sim 0 $\\
				$\Delta m^2_{\text{atm}}~ [{\rm eV}^2]$ & $2.511\times10^{-3}$& $-$& $-$ &$2.586 \times 10^{-3}$ & $\sim 0$ & $2.572 \times 10^{-3}$ & $\sim 0$ \\
				$\sin^2 \theta _{12}$ & $0.303$ & $-$ & $-$&$0.303$ & $\sim 0$ & $0.303$ & $\sim 0$\\
				$\sin^2 \theta _{23}$ & $0.572$   & $-$ & $-$& $0.558$ & $-0.2 $ & $0.571$ & $\sim 0$  \\
				$\sin^2 \theta _{13}$ & $0.02203$ & $-$  & $-$ &$0.02194$& $\sim 0$ & $0.02201$ & $\sim 0$ \\
                $\delta_{\rm{MNS}}\,[^\circ]$ & $197$ &$-$ & $-$ & $192$ & $-0.2$ & $197$ & $\sim 0$\\
    \hline
                $\chi^2_{\rm min}$    & & & $2.0$ & & $4.0$ & & $3.1$\\
    \hline
                $M_T$ [GeV]& &$10^{12}$ & & $7.7 \times 10^{17}$& & $10^{13} $ &\\
                $M_{N_1}$ [GeV]& &$4.7\times 10^{18}$ & & $ 4.1 \times 10^{16}$ & & $5.6 \times 10^{12}$&\\
                $M_{N_2}$ [GeV] & & $-$& & $2.3 \times 10^{12}$& & $6.9 \times 10^{17}$ &\\
                $M_{N_3}$ [GeV] & & $-$ & & $-$ & &$1.2\times 10^{13}$ & \\ 
				\hline
				\hline 
			\end{tabular}
		\end{math}
	\end{center}
	\caption{The benchmark best-fit solutions obtained for three example cases as discussed in the text. $O_{\rm exp}$ denote the extrapolated values of the underlying observables at $\mu=10^{16}$ GeV. The reproduced values through $\chi^2$ minimisation are listed under $O_{\rm th}$ and corresponding pulls are given for each solution. The optimized values of the masses of leptoquark scalar and RH neutrinos are listed at the bottom of the table.} 
	\label{tab:tab1} 
\end{table*}
%%%%%%%%%%%%%

Although, the RH neutrino is introduced to reproduce the viable charged fermion mass spectrum, its mass and couplings are not constrained from the requirement of the light neutrino masses and mixing parameters. To account for both the solar and atmospheric neutrino mass scales, one needs at least two RH neutrinos in the minimal realization. The light neutrino masses are then generated through the usual type I seesaw mechanism:
\be \label{seesaw}
M_\nu = - v^2\, Y_\nu M_N^{-1} Y_\nu^T\,. \ee
Here, $M_\nu$ is $3 \times 3$ light neutrino mass matrix while $M_N$ is $2\times2$ heavy neutrino mass matrix. $Y_\nu$ is $3 \times 2$ matrix which can be computed using Eqs. (\ref{dY_gen},\ref{dY},\ref{K}). The above leads to one massless light neutrino.

We extend the $\chi^2$ function to include the solar and atmospheric squared mass differences, three mixing angles and a Dirac CP phase to assess if Eq. (\ref{seesaw}) along with Eqs. (\ref{dY_gen},\ref{dY},\ref{K}) can provide a realistic spectrum of quarks and leptons. For the input values of neutrino observables, we use the results of the latest fit from \cite{Esteban:2020cvm} and set $\pm 10 \%$ uncertainty as earlier. The RGE effects in neutrino data are neglected as they are known to be small \cite{Chankowski:1993tx,Babu:1993qv,Antusch:2005gp,Mei:2005qp} and within the set uncertainty for normal hierarchy in the neutrino masses which is the case considered here.

The result of $\chi^2$ minimization for this case is shown in Table \ref{tab:tab1} as Solution II and the optimized values of parameters are listed in the Appendix \ref{app:sol}. As it can be seen, we find a very good agreement with all the fermion masses and mixing parameters with $\chi^2_{\rm min} =4$. The resulting values of $M_{N_1}$ and $M_{N_2}$ are smaller than $M_{\rm GUT}$ which requires $M_T > 10^{17.2}$ GeV as anticipated from Fig. \ref{fig2}.

As a simple extension of the possibilities discussed above, it is straightforward to anticipate a case in which there are more than two RH neutrinos present. At least one of them is strongly coupled with the SM fermions and has a mass greater than $M_{\rm GUT}$. It leads to the required threshold corrections for a viable charged fermion spectrum, however its contribution to the neutrino masses is sub-dominant. The other RH neutrinos have sub-GUT scale masses and can lead to a realistic light neutrino spectrum without significantly altering the threshold corrections. This scenario is exemplified by Solution III in Table \ref{tab:tab1}. In this case, $N_3$, with $M_{N_2} > M_{\rm{GUT}}$, couples to the SM leptons with large couplings and gives the required threshold corrections to down-type quark sector. Notably, in this context, it is evident that the colour triplet scalar need not approach Planck-scale values to fulfil its role.

\section{Conclusion}
\label{sec:concl}
This article demonstrates that the seemingly unviable relationship, $Y_d = Y_e^T$, predicted by the simplest and most minimal Yukawa sector of non-supersymmetric $SU(5)$ GUT, can be rendered viable when accounting for 1-loop corrections to the tree-level matching conditions. This is accomplished by introducing one or more copies of fermion singlets. While they do not alter the tree-level matching conditions at the scale of unification, they can yield significant corrections at the 1-loop level through their direct Yukawa interactions with the down-type quarks and the colour triplet scalar. Sizeable non-degeneracy among the singlet fermions, colour triplet scalar, and leptoquark vector can thus impart large enough threshold corrections ensuring the compatibility of the minimal Yukawa sector with the effective SM description.

Our quantitative analysis reveals that achieving a realistic spectrum for the charged fermion Yukawa couplings and quark mixing necessitates either a significantly larger mass for the colour triplet scalar ($M_T \gg M_X$) or vastly higher masses for the RH neutrinos ($M_{N_\alpha} \gg M_X$), under the assumption that the mass of the leptoquark gauge boson ($M_X$) defines the unification scale. The latter possibility is disfavoured if the same fermion singlets are expected to generate a viable light neutrino spectrum through the conventional type I seesaw mechanism. Nonetheless, the scenario of $M_{N_\alpha} \gg M_X$ remains a plausible option if neutrinos acquire their masses through other means. This also includes type I seesaw mechanism with additional copies of RH neutrinos with sub-GUT scale masses and comparatively smaller couplings with the SM leptons.

It is noteworthy that the inclusion of quantum corrections can substantially alter the conclusions regarding the minimal Yukawa sector within the framework of an underlying grand unified theory. These findings provide motivation for conducting analogous investigations in the context of supersymmetric variants of $SU(5)$\footnote{Radiative corrections to $Y_d = Y_e^T$ arising from the superpartners of the SM fields in the loop have been considered in \cite{Diaz-Cruz:2000nvf,Enkhbat:2009jt}. This, however, requires non-minimal supersymmetry breaking trilinear terms.}, as well as both the ordinary and supersymmetric versions of $SO(10)$ GUTs, which feature more diverse particle spectra for threshold corrections and, simultaneously, more stringent symmetries that engage in intricate interplays.

\section*{Acknowledgements}
We acknowledge illuminating discussions with Charanjit Singh Aulakh and Anjan S. Joshipura. This work is partially supported under the MATRICS project (MTR/2021/000049) funded by the Science \& Engineering Research Board (SERB), Department of Science and Technology (DST), Government of India. KMP also acknowledges support from the ICTP through the Associates Programme (2023-2028) where part of this work was carried out.

\bibliography{references}

%merlin.mbs apsrev4-1.bst 2010-07-25 4.21a (PWD, AO, DPC) hacked
%Control: key (0)
%Control: author (72) initials jnrlst
%Control: editor formatted (1) identically to author
%Control: production of article title (-1) disabled
%Control: page (0) single
%Control: year (1) truncated
%Control: production of eprint (0) enabled
\begin{thebibliography}{59}%
\makeatletter
\providecommand \@ifxundefined [1]{%
 \@ifx{#1\undefined}
}%
\providecommand \@ifnum [1]{%
 \ifnum #1\expandafter \@firstoftwo
 \else \expandafter \@secondoftwo
 \fi
}%
\providecommand \@ifx [1]{%
 \ifx #1\expandafter \@firstoftwo
 \else \expandafter \@secondoftwo
 \fi
}%
\providecommand \natexlab [1]{#1}%
\providecommand \enquote  [1]{``#1''}%
\providecommand \bibnamefont  [1]{#1}%
\providecommand \bibfnamefont [1]{#1}%
\providecommand \citenamefont [1]{#1}%
\providecommand \href@noop [0]{\@secondoftwo}%
\providecommand \href [0]{\begingroup \@sanitize@url \@href}%
\providecommand \@href[1]{\@@startlink{#1}\@@href}%
\providecommand \@@href[1]{\endgroup#1\@@endlink}%
\providecommand \@sanitize@url [0]{\catcode `\\12\catcode `\$12\catcode
  `\&12\catcode `\#12\catcode `\^12\catcode `\_12\catcode `\%12\relax}%
\providecommand \@@startlink[1]{}%
\providecommand \@@endlink[0]{}%
\providecommand \url  [0]{\begingroup\@sanitize@url \@url }%
\providecommand \@url [1]{\endgroup\@href {#1}{\urlprefix }}%
\providecommand \urlprefix  [0]{URL }%
\providecommand \Eprint [0]{\href }%
\providecommand \doibase [0]{http://dx.doi.org/}%
\providecommand \selectlanguage [0]{\@gobble}%
\providecommand \bibinfo  [0]{\@secondoftwo}%
\providecommand \bibfield  [0]{\@secondoftwo}%
\providecommand \translation [1]{[#1]}%
\providecommand \BibitemOpen [0]{}%
\providecommand \bibitemStop [0]{}%
\providecommand \bibitemNoStop [0]{.\EOS\space}%
\providecommand \EOS [0]{\spacefactor3000\relax}%
\providecommand \BibitemShut  [1]{\csname bibitem#1\endcsname}%
\let\auto@bib@innerbib\@empty
%</preamble>
\bibitem [{\citenamefont {Georgi}\ and\ \citenamefont
  {Glashow}(1974)}]{Georgi:1974sy}%
  \BibitemOpen
  \bibfield  {author} {\bibinfo {author} {\bibfnamefont {H.}~\bibnamefont
  {Georgi}}\ and\ \bibinfo {author} {\bibfnamefont {S.~L.}\ \bibnamefont
  {Glashow}},\ }\href {\doibase 10.1103/PhysRevLett.32.438} {\bibfield
  {journal} {\bibinfo  {journal} {Phys. Rev. Lett.}\ }\textbf {\bibinfo
  {volume} {32}},\ \bibinfo {pages} {438} (\bibinfo {year} {1974})}\BibitemShut
  {NoStop}%
\bibitem [{\citenamefont {Fritzsch}\ and\ \citenamefont
  {Minkowski}(1975)}]{Fritzsch:1974nn}%
  \BibitemOpen
  \bibfield  {author} {\bibinfo {author} {\bibfnamefont {H.}~\bibnamefont
  {Fritzsch}}\ and\ \bibinfo {author} {\bibfnamefont {P.}~\bibnamefont
  {Minkowski}},\ }\href {\doibase 10.1016/0003-4916(75)90211-0} {\bibfield
  {journal} {\bibinfo  {journal} {Annals Phys.}\ }\textbf {\bibinfo {volume}
  {93}},\ \bibinfo {pages} {193} (\bibinfo {year} {1975})}\BibitemShut
  {NoStop}%
\bibitem [{\citenamefont {Pati}\ and\ \citenamefont
  {Salam}(1974)}]{Pati:1974yy}%
  \BibitemOpen
  \bibfield  {author} {\bibinfo {author} {\bibfnamefont {J.~C.}\ \bibnamefont
  {Pati}}\ and\ \bibinfo {author} {\bibfnamefont {A.}~\bibnamefont {Salam}},\
  }\href {\doibase 10.1103/PhysRevD.10.275} {\bibfield  {journal} {\bibinfo
  {journal} {Phys. Rev. D}\ }\textbf {\bibinfo {volume} {10}},\ \bibinfo
  {pages} {275} (\bibinfo {year} {1974})},\ \bibinfo {note} {[Erratum:
  Phys.Rev.D 11, 703--703 (1975)]}\BibitemShut {NoStop}%
\bibitem [{\citenamefont {Buras}\ \emph {et~al.}(1978)\citenamefont {Buras},
  \citenamefont {Ellis}, \citenamefont {Gaillard},\ and\ \citenamefont
  {Nanopoulos}}]{Buras:1977yy}%
  \BibitemOpen
  \bibfield  {author} {\bibinfo {author} {\bibfnamefont {A.~J.}\ \bibnamefont
  {Buras}}, \bibinfo {author} {\bibfnamefont {J.~R.}\ \bibnamefont {Ellis}},
  \bibinfo {author} {\bibfnamefont {M.~K.}\ \bibnamefont {Gaillard}}, \ and\
  \bibinfo {author} {\bibfnamefont {D.~V.}\ \bibnamefont {Nanopoulos}},\ }\href
  {\doibase 10.1016/0550-3213(78)90214-6} {\bibfield  {journal} {\bibinfo
  {journal} {Nucl. Phys. B}\ }\textbf {\bibinfo {volume} {135}},\ \bibinfo
  {pages} {66} (\bibinfo {year} {1978})}\BibitemShut {NoStop}%
\bibitem [{\citenamefont {Langacker}(1981)}]{Langacker:1980js}%
  \BibitemOpen
  \bibfield  {author} {\bibinfo {author} {\bibfnamefont {P.}~\bibnamefont
  {Langacker}},\ }\href {\doibase 10.1016/0370-1573(81)90059-4} {\bibfield
  {journal} {\bibinfo  {journal} {Phys. Rept.}\ }\textbf {\bibinfo {volume}
  {72}},\ \bibinfo {pages} {185} (\bibinfo {year} {1981})}\BibitemShut
  {NoStop}%
\bibitem [{\citenamefont {Georgi}\ and\ \citenamefont
  {Jarlskog}(1979)}]{Georgi:1979df}%
  \BibitemOpen
  \bibfield  {author} {\bibinfo {author} {\bibfnamefont {H.}~\bibnamefont
  {Georgi}}\ and\ \bibinfo {author} {\bibfnamefont {C.}~\bibnamefont
  {Jarlskog}},\ }\href {\doibase 10.1016/0370-2693(79)90842-6} {\bibfield
  {journal} {\bibinfo  {journal} {Phys. Lett. B}\ }\textbf {\bibinfo {volume}
  {86}},\ \bibinfo {pages} {297} (\bibinfo {year} {1979})}\BibitemShut
  {NoStop}%
\bibitem [{\citenamefont {Barbieri}\ \emph {et~al.}(1981)\citenamefont
  {Barbieri}, \citenamefont {Nanopoulos},\ and\ \citenamefont
  {Wyler}}]{Barbieri:1981yw}%
  \BibitemOpen
  \bibfield  {author} {\bibinfo {author} {\bibfnamefont {R.}~\bibnamefont
  {Barbieri}}, \bibinfo {author} {\bibfnamefont {D.~V.}\ \bibnamefont
  {Nanopoulos}}, \ and\ \bibinfo {author} {\bibfnamefont {D.}~\bibnamefont
  {Wyler}},\ }\href {\doibase 10.1016/0370-2693(81)90076-9} {\bibfield
  {journal} {\bibinfo  {journal} {Phys. Lett. B}\ }\textbf {\bibinfo {volume}
  {103}},\ \bibinfo {pages} {433} (\bibinfo {year} {1981})}\BibitemShut
  {NoStop}%
\bibitem [{\citenamefont {Giveon}\ \emph {et~al.}(1991)\citenamefont {Giveon},
  \citenamefont {Hall},\ and\ \citenamefont {Sarid}}]{Giveon:1991zm}%
  \BibitemOpen
  \bibfield  {author} {\bibinfo {author} {\bibfnamefont {A.}~\bibnamefont
  {Giveon}}, \bibinfo {author} {\bibfnamefont {L.~J.}\ \bibnamefont {Hall}}, \
  and\ \bibinfo {author} {\bibfnamefont {U.}~\bibnamefont {Sarid}},\ }\href
  {\doibase 10.1016/0370-2693(91)91289-8} {\bibfield  {journal} {\bibinfo
  {journal} {Phys. Lett. B}\ }\textbf {\bibinfo {volume} {271}},\ \bibinfo
  {pages} {138} (\bibinfo {year} {1991})}\BibitemShut {NoStop}%
\bibitem [{\citenamefont {Dorsner}\ and\ \citenamefont
  {Fileviez~Perez}(2006)}]{Dorsner:2006dj}%
  \BibitemOpen
  \bibfield  {author} {\bibinfo {author} {\bibfnamefont {I.}~\bibnamefont
  {Dorsner}}\ and\ \bibinfo {author} {\bibfnamefont {P.}~\bibnamefont
  {Fileviez~Perez}},\ }\href {\doibase 10.1016/j.physletb.2006.09.034}
  {\bibfield  {journal} {\bibinfo  {journal} {Phys. Lett. B}\ }\textbf
  {\bibinfo {volume} {642}},\ \bibinfo {pages} {248} (\bibinfo {year}
  {2006})},\ \Eprint {http://arxiv.org/abs/hep-ph/0606062}
  {arXiv:hep-ph/0606062} \BibitemShut {NoStop}%
\bibitem [{\citenamefont {Fileviez~Perez}(2007)}]{FileviezPerez:2007bcw}%
  \BibitemOpen
  \bibfield  {author} {\bibinfo {author} {\bibfnamefont {P.}~\bibnamefont
  {Fileviez~Perez}},\ }\href {\doibase 10.1016/j.physletb.2007.07.075}
  {\bibfield  {journal} {\bibinfo  {journal} {Phys. Lett. B}\ }\textbf
  {\bibinfo {volume} {654}},\ \bibinfo {pages} {189} (\bibinfo {year}
  {2007})},\ \Eprint {http://arxiv.org/abs/hep-ph/0702287}
  {arXiv:hep-ph/0702287} \BibitemShut {NoStop}%
\bibitem [{\citenamefont {Goto}\ \emph {et~al.}(2023)\citenamefont {Goto},
  \citenamefont {Mishima},\ and\ \citenamefont {Shindou}}]{Goto:2023qch}%
  \BibitemOpen
  \bibfield  {author} {\bibinfo {author} {\bibfnamefont {T.}~\bibnamefont
  {Goto}}, \bibinfo {author} {\bibfnamefont {S.}~\bibnamefont {Mishima}}, \
  and\ \bibinfo {author} {\bibfnamefont {T.}~\bibnamefont {Shindou}},\
  }\href@noop {} {\  (\bibinfo {year} {2023})},\ \Eprint
  {http://arxiv.org/abs/2308.13329} {arXiv:2308.13329 [hep-ph]} \BibitemShut
  {NoStop}%
\bibitem [{\citenamefont {Ellis}\ and\ \citenamefont
  {Gaillard}(1979)}]{Ellis:1979fg}%
  \BibitemOpen
  \bibfield  {author} {\bibinfo {author} {\bibfnamefont {J.~R.}\ \bibnamefont
  {Ellis}}\ and\ \bibinfo {author} {\bibfnamefont {M.~K.}\ \bibnamefont
  {Gaillard}},\ }\href {\doibase 10.1016/0370-2693(79)90476-3} {\bibfield
  {journal} {\bibinfo  {journal} {Phys. Lett. B}\ }\textbf {\bibinfo {volume}
  {88}},\ \bibinfo {pages} {315} (\bibinfo {year} {1979})}\BibitemShut
  {NoStop}%
\bibitem [{\citenamefont {Berezinsky}\ and\ \citenamefont
  {Smirnov}(1984)}]{Berezinsky:1983va}%
  \BibitemOpen
  \bibfield  {author} {\bibinfo {author} {\bibfnamefont {V.~S.}\ \bibnamefont
  {Berezinsky}}\ and\ \bibinfo {author} {\bibfnamefont {A.~Y.}\ \bibnamefont
  {Smirnov}},\ }\href {\doibase 10.1016/0370-2693(84)91044-X} {\bibfield
  {journal} {\bibinfo  {journal} {Phys. Lett. B}\ }\textbf {\bibinfo {volume}
  {140}},\ \bibinfo {pages} {49} (\bibinfo {year} {1984})}\BibitemShut
  {NoStop}%
\bibitem [{\citenamefont {Altarelli}\ \emph {et~al.}(2000)\citenamefont
  {Altarelli}, \citenamefont {Feruglio},\ and\ \citenamefont
  {Masina}}]{Altarelli:2000fu}%
  \BibitemOpen
  \bibfield  {author} {\bibinfo {author} {\bibfnamefont {G.}~\bibnamefont
  {Altarelli}}, \bibinfo {author} {\bibfnamefont {F.}~\bibnamefont {Feruglio}},
  \ and\ \bibinfo {author} {\bibfnamefont {I.}~\bibnamefont {Masina}},\ }\href
  {\doibase 10.1088/1126-6708/2000/11/040} {\bibfield  {journal} {\bibinfo
  {journal} {JHEP}\ }\textbf {\bibinfo {volume} {11}},\ \bibinfo {pages} {040}
  (\bibinfo {year} {2000})},\ \Eprint {http://arxiv.org/abs/hep-ph/0007254}
  {arXiv:hep-ph/0007254} \BibitemShut {NoStop}%
\bibitem [{\citenamefont {Emmanuel-Costa}\ and\ \citenamefont
  {Wiesenfeldt}(2003)}]{Emmanuel-Costa:2003szk}%
  \BibitemOpen
  \bibfield  {author} {\bibinfo {author} {\bibfnamefont {D.}~\bibnamefont
  {Emmanuel-Costa}}\ and\ \bibinfo {author} {\bibfnamefont {S.}~\bibnamefont
  {Wiesenfeldt}},\ }\href {\doibase 10.1016/S0550-3213(03)00301-8} {\bibfield
  {journal} {\bibinfo  {journal} {Nucl. Phys. B}\ }\textbf {\bibinfo {volume}
  {661}},\ \bibinfo {pages} {62} (\bibinfo {year} {2003})},\ \Eprint
  {http://arxiv.org/abs/hep-ph/0302272} {arXiv:hep-ph/0302272} \BibitemShut
  {NoStop}%
\bibitem [{\citenamefont {Dorsner}\ \emph {et~al.}(2007)\citenamefont
  {Dorsner}, \citenamefont {Fileviez~Perez},\ and\ \citenamefont
  {Rodrigo}}]{Dorsner:2006hw}%
  \BibitemOpen
  \bibfield  {author} {\bibinfo {author} {\bibfnamefont {I.}~\bibnamefont
  {Dorsner}}, \bibinfo {author} {\bibfnamefont {P.}~\bibnamefont
  {Fileviez~Perez}}, \ and\ \bibinfo {author} {\bibfnamefont {G.}~\bibnamefont
  {Rodrigo}},\ }\href {\doibase 10.1103/PhysRevD.75.125007} {\bibfield
  {journal} {\bibinfo  {journal} {Phys. Rev. D}\ }\textbf {\bibinfo {volume}
  {75}},\ \bibinfo {pages} {125007} (\bibinfo {year} {2007})},\ \Eprint
  {http://arxiv.org/abs/hep-ph/0607208} {arXiv:hep-ph/0607208} \BibitemShut
  {NoStop}%
\bibitem [{\citenamefont {Antusch}\ and\ \citenamefont
  {Hinze}(2022)}]{Antusch:2021yqe}%
  \BibitemOpen
  \bibfield  {author} {\bibinfo {author} {\bibfnamefont {S.}~\bibnamefont
  {Antusch}}\ and\ \bibinfo {author} {\bibfnamefont {K.}~\bibnamefont
  {Hinze}},\ }\href {\doibase 10.1016/j.nuclphysb.2022.115719} {\bibfield
  {journal} {\bibinfo  {journal} {Nucl. Phys. B}\ }\textbf {\bibinfo {volume}
  {976}},\ \bibinfo {pages} {115719} (\bibinfo {year} {2022})},\ \Eprint
  {http://arxiv.org/abs/2108.08080} {arXiv:2108.08080 [hep-ph]} \BibitemShut
  {NoStop}%
\bibitem [{\citenamefont {Antusch}\ \emph
  {et~al.}(2023{\natexlab{a}})\citenamefont {Antusch}, \citenamefont {Hinze},\
  and\ \citenamefont {Saad}}]{Antusch:2022afk}%
  \BibitemOpen
  \bibfield  {author} {\bibinfo {author} {\bibfnamefont {S.}~\bibnamefont
  {Antusch}}, \bibinfo {author} {\bibfnamefont {K.}~\bibnamefont {Hinze}}, \
  and\ \bibinfo {author} {\bibfnamefont {S.}~\bibnamefont {Saad}},\ }\href
  {\doibase 10.1016/j.nuclphysb.2022.116049} {\bibfield  {journal} {\bibinfo
  {journal} {Nucl. Phys. B}\ }\textbf {\bibinfo {volume} {986}},\ \bibinfo
  {pages} {116049} (\bibinfo {year} {2023}{\natexlab{a}})},\ \Eprint
  {http://arxiv.org/abs/2205.01120} {arXiv:2205.01120 [hep-ph]} \BibitemShut
  {NoStop}%
\bibitem [{\citenamefont {Hempfling}(1994)}]{Hempfling:1993kv}%
  \BibitemOpen
  \bibfield  {author} {\bibinfo {author} {\bibfnamefont {R.}~\bibnamefont
  {Hempfling}},\ }\href {\doibase 10.1103/PhysRevD.49.6168} {\bibfield
  {journal} {\bibinfo  {journal} {Phys. Rev. D}\ }\textbf {\bibinfo {volume}
  {49}},\ \bibinfo {pages} {6168} (\bibinfo {year} {1994})}\BibitemShut
  {NoStop}%
\bibitem [{\citenamefont {Shafi}\ and\ \citenamefont
  {Tavartkiladze}(1999)}]{Shafi:1999rm}%
  \BibitemOpen
  \bibfield  {author} {\bibinfo {author} {\bibfnamefont {Q.}~\bibnamefont
  {Shafi}}\ and\ \bibinfo {author} {\bibfnamefont {Z.}~\bibnamefont
  {Tavartkiladze}},\ }\href {\doibase 10.1016/S0370-2693(99)00185-9} {\bibfield
   {journal} {\bibinfo  {journal} {Phys. Lett. B}\ }\textbf {\bibinfo {volume}
  {451}},\ \bibinfo {pages} {129} (\bibinfo {year} {1999})},\ \Eprint
  {http://arxiv.org/abs/hep-ph/9901243} {arXiv:hep-ph/9901243} \BibitemShut
  {NoStop}%
\bibitem [{\citenamefont {Barr}\ and\ \citenamefont
  {Dorsner}(2003)}]{Barr:2003zx}%
  \BibitemOpen
  \bibfield  {author} {\bibinfo {author} {\bibfnamefont {S.~M.}\ \bibnamefont
  {Barr}}\ and\ \bibinfo {author} {\bibfnamefont {I.}~\bibnamefont {Dorsner}},\
  }\href {\doibase 10.1016/S0370-2693(03)00772-X} {\bibfield  {journal}
  {\bibinfo  {journal} {Phys. Lett. B}\ }\textbf {\bibinfo {volume} {566}},\
  \bibinfo {pages} {125} (\bibinfo {year} {2003})},\ \Eprint
  {http://arxiv.org/abs/hep-ph/0305090} {arXiv:hep-ph/0305090} \BibitemShut
  {NoStop}%
\bibitem [{\citenamefont {Oshimo}(2009)}]{Oshimo:2009ia}%
  \BibitemOpen
  \bibfield  {author} {\bibinfo {author} {\bibfnamefont {N.}~\bibnamefont
  {Oshimo}},\ }\href {\doibase 10.1103/PhysRevD.80.075011} {\bibfield
  {journal} {\bibinfo  {journal} {Phys. Rev. D}\ }\textbf {\bibinfo {volume}
  {80}},\ \bibinfo {pages} {075011} (\bibinfo {year} {2009})},\ \Eprint
  {http://arxiv.org/abs/0907.3400} {arXiv:0907.3400 [hep-ph]} \BibitemShut
  {NoStop}%
\bibitem [{\citenamefont {Babu}\ \emph {et~al.}(2012)\citenamefont {Babu},
  \citenamefont {Bajc},\ and\ \citenamefont {Tavartkiladze}}]{Babu:2012pb}%
  \BibitemOpen
  \bibfield  {author} {\bibinfo {author} {\bibfnamefont {K.~S.}\ \bibnamefont
  {Babu}}, \bibinfo {author} {\bibfnamefont {B.}~\bibnamefont {Bajc}}, \ and\
  \bibinfo {author} {\bibfnamefont {Z.}~\bibnamefont {Tavartkiladze}},\ }\href
  {\doibase 10.1103/PhysRevD.86.075005} {\bibfield  {journal} {\bibinfo
  {journal} {Phys. Rev. D}\ }\textbf {\bibinfo {volume} {86}},\ \bibinfo
  {pages} {075005} (\bibinfo {year} {2012})},\ \Eprint
  {http://arxiv.org/abs/1207.6388} {arXiv:1207.6388 [hep-ph]} \BibitemShut
  {NoStop}%
\bibitem [{\citenamefont {Dorsner}\ \emph {et~al.}(2014)\citenamefont
  {Dorsner}, \citenamefont {Fajfer},\ and\ \citenamefont
  {Mustac}}]{Dorsner:2014wva}%
  \BibitemOpen
  \bibfield  {author} {\bibinfo {author} {\bibfnamefont {I.}~\bibnamefont
  {Dorsner}}, \bibinfo {author} {\bibfnamefont {S.}~\bibnamefont {Fajfer}}, \
  and\ \bibinfo {author} {\bibfnamefont {I.}~\bibnamefont {Mustac}},\ }\href
  {\doibase 10.1103/PhysRevD.89.115004} {\bibfield  {journal} {\bibinfo
  {journal} {Phys. Rev. D}\ }\textbf {\bibinfo {volume} {89}},\ \bibinfo
  {pages} {115004} (\bibinfo {year} {2014})},\ \Eprint
  {http://arxiv.org/abs/1401.6870} {arXiv:1401.6870 [hep-ph]} \BibitemShut
  {NoStop}%
\bibitem [{\citenamefont {Fileviez~P\'erez}\ \emph {et~al.}(2018)\citenamefont
  {Fileviez~P\'erez}, \citenamefont {Gross},\ and\ \citenamefont
  {Murgui}}]{FileviezPerez:2018dyf}%
  \BibitemOpen
  \bibfield  {author} {\bibinfo {author} {\bibfnamefont {P.}~\bibnamefont
  {Fileviez~P\'erez}}, \bibinfo {author} {\bibfnamefont {A.}~\bibnamefont
  {Gross}}, \ and\ \bibinfo {author} {\bibfnamefont {C.}~\bibnamefont
  {Murgui}},\ }\href {\doibase 10.1103/PhysRevD.98.035032} {\bibfield
  {journal} {\bibinfo  {journal} {Phys. Rev. D}\ }\textbf {\bibinfo {volume}
  {98}},\ \bibinfo {pages} {035032} (\bibinfo {year} {2018})},\ \Eprint
  {http://arxiv.org/abs/1804.07831} {arXiv:1804.07831 [hep-ph]} \BibitemShut
  {NoStop}%
\bibitem [{\citenamefont {Antusch}\ \emph
  {et~al.}(2023{\natexlab{b}})\citenamefont {Antusch}, \citenamefont {Hinze},
  \citenamefont {Saad},\ and\ \citenamefont {Steiner}}]{Antusch:2023mxx}%
  \BibitemOpen
  \bibfield  {author} {\bibinfo {author} {\bibfnamefont {S.}~\bibnamefont
  {Antusch}}, \bibinfo {author} {\bibfnamefont {K.}~\bibnamefont {Hinze}},
  \bibinfo {author} {\bibfnamefont {S.}~\bibnamefont {Saad}}, \ and\ \bibinfo
  {author} {\bibfnamefont {J.}~\bibnamefont {Steiner}},\ }\href@noop {} {\
  (\bibinfo {year} {2023}{\natexlab{b}})},\ \Eprint
  {http://arxiv.org/abs/2308.11705} {arXiv:2308.11705 [hep-ph]} \BibitemShut
  {NoStop}%
\bibitem [{\citenamefont {Antusch}\ \emph
  {et~al.}(2023{\natexlab{c}})\citenamefont {Antusch}, \citenamefont {Hinze},\
  and\ \citenamefont {Saad}}]{Antusch:2023kli}%
  \BibitemOpen
  \bibfield  {author} {\bibinfo {author} {\bibfnamefont {S.}~\bibnamefont
  {Antusch}}, \bibinfo {author} {\bibfnamefont {K.}~\bibnamefont {Hinze}}, \
  and\ \bibinfo {author} {\bibfnamefont {S.}~\bibnamefont {Saad}},\ }\href
  {\doibase 10.1016/j.nuclphysb.2023.116195} {\bibfield  {journal} {\bibinfo
  {journal} {Nucl. Phys. B}\ }\textbf {\bibinfo {volume} {991}},\ \bibinfo
  {pages} {116195} (\bibinfo {year} {2023}{\natexlab{c}})},\ \Eprint
  {http://arxiv.org/abs/2301.03601} {arXiv:2301.03601 [hep-ph]} \BibitemShut
  {NoStop}%
\bibitem [{\citenamefont {Antusch}\ \emph
  {et~al.}(2023{\natexlab{d}})\citenamefont {Antusch}, \citenamefont {Hinze},\
  and\ \citenamefont {Saad}}]{Antusch:2023mqe}%
  \BibitemOpen
  \bibfield  {author} {\bibinfo {author} {\bibfnamefont {S.}~\bibnamefont
  {Antusch}}, \bibinfo {author} {\bibfnamefont {K.}~\bibnamefont {Hinze}}, \
  and\ \bibinfo {author} {\bibfnamefont {S.}~\bibnamefont {Saad}},\ }\href@noop
  {} {\  (\bibinfo {year} {2023}{\natexlab{d}})},\ \Eprint
  {http://arxiv.org/abs/2308.08585} {arXiv:2308.08585 [hep-ph]} \BibitemShut
  {NoStop}%
\bibitem [{\citenamefont {Aulakh}\ and\ \citenamefont
  {Garg}(2006)}]{Aulakh:2006hs}%
  \BibitemOpen
  \bibfield  {author} {\bibinfo {author} {\bibfnamefont {C.~S.}\ \bibnamefont
  {Aulakh}}\ and\ \bibinfo {author} {\bibfnamefont {S.~K.}\ \bibnamefont
  {Garg}},\ }\href@noop {} {\  (\bibinfo {year} {2006})},\ \Eprint
  {http://arxiv.org/abs/hep-ph/0612021} {arXiv:hep-ph/0612021} \BibitemShut
  {NoStop}%
\bibitem [{\citenamefont {Aulakh}\ and\ \citenamefont
  {Garg}(2012)}]{Aulakh:2008sn}%
  \BibitemOpen
  \bibfield  {author} {\bibinfo {author} {\bibfnamefont {C.~S.}\ \bibnamefont
  {Aulakh}}\ and\ \bibinfo {author} {\bibfnamefont {S.~K.}\ \bibnamefont
  {Garg}},\ }\href {\doibase 10.1016/j.nuclphysb.2011.12.003} {\bibfield
  {journal} {\bibinfo  {journal} {Nucl. Phys. B}\ }\textbf {\bibinfo {volume}
  {857}},\ \bibinfo {pages} {101} (\bibinfo {year} {2012})},\ \Eprint
  {http://arxiv.org/abs/0807.0917} {arXiv:0807.0917 [hep-ph]} \BibitemShut
  {NoStop}%
\bibitem [{\citenamefont {Aulakh}\ \emph {et~al.}(2014)\citenamefont {Aulakh},
  \citenamefont {Garg},\ and\ \citenamefont {Khosa}}]{Aulakh:2013lxa}%
  \BibitemOpen
  \bibfield  {author} {\bibinfo {author} {\bibfnamefont {C.~S.}\ \bibnamefont
  {Aulakh}}, \bibinfo {author} {\bibfnamefont {I.}~\bibnamefont {Garg}}, \ and\
  \bibinfo {author} {\bibfnamefont {C.~K.}\ \bibnamefont {Khosa}},\ }\href
  {\doibase 10.1016/j.nuclphysb.2014.03.003} {\bibfield  {journal} {\bibinfo
  {journal} {Nucl. Phys. B}\ }\textbf {\bibinfo {volume} {882}},\ \bibinfo
  {pages} {397} (\bibinfo {year} {2014})},\ \Eprint
  {http://arxiv.org/abs/1311.6100} {arXiv:1311.6100 [hep-ph]} \BibitemShut
  {NoStop}%
\bibitem [{\citenamefont {Patel}\ and\ \citenamefont
  {Shukla}(2022)}]{Patel:2022wya}%
  \BibitemOpen
  \bibfield  {author} {\bibinfo {author} {\bibfnamefont {K.~M.}\ \bibnamefont
  {Patel}}\ and\ \bibinfo {author} {\bibfnamefont {S.~K.}\ \bibnamefont
  {Shukla}},\ }\href {\doibase 10.1007/JHEP08(2022)042} {\bibfield  {journal}
  {\bibinfo  {journal} {JHEP}\ }\textbf {\bibinfo {volume} {08}},\ \bibinfo
  {pages} {042} (\bibinfo {year} {2022})},\ \Eprint
  {http://arxiv.org/abs/2203.07748} {arXiv:2203.07748 [hep-ph]} \BibitemShut
  {NoStop}%
\bibitem [{\citenamefont {Wright}(1994)}]{Wright:1994qb}%
  \BibitemOpen
  \bibfield  {author} {\bibinfo {author} {\bibfnamefont {B.~D.}\ \bibnamefont
  {Wright}},\ }\href@noop {} {\  (\bibinfo {year} {1994})},\ \Eprint
  {http://arxiv.org/abs/hep-ph/9404217} {arXiv:hep-ph/9404217} \BibitemShut
  {NoStop}%
\bibitem [{\citenamefont {Grisaru}\ \emph {et~al.}(1979)\citenamefont
  {Grisaru}, \citenamefont {Siegel},\ and\ \citenamefont
  {Rocek}}]{Grisaru:1979wc}%
  \BibitemOpen
  \bibfield  {author} {\bibinfo {author} {\bibfnamefont {M.~T.}\ \bibnamefont
  {Grisaru}}, \bibinfo {author} {\bibfnamefont {W.}~\bibnamefont {Siegel}}, \
  and\ \bibinfo {author} {\bibfnamefont {M.}~\bibnamefont {Rocek}},\ }\href
  {\doibase 10.1016/0550-3213(79)90344-4} {\bibfield  {journal} {\bibinfo
  {journal} {Nucl. Phys. B}\ }\textbf {\bibinfo {volume} {159}},\ \bibinfo
  {pages} {429} (\bibinfo {year} {1979})}\BibitemShut {NoStop}%
\bibitem [{\citenamefont {Seiberg}(1993)}]{Seiberg:1993vc}%
  \BibitemOpen
  \bibfield  {author} {\bibinfo {author} {\bibfnamefont {N.}~\bibnamefont
  {Seiberg}},\ }\href {\doibase 10.1016/0370-2693(93)91541-T} {\bibfield
  {journal} {\bibinfo  {journal} {Phys. Lett. B}\ }\textbf {\bibinfo {volume}
  {318}},\ \bibinfo {pages} {469} (\bibinfo {year} {1993})},\ \Eprint
  {http://arxiv.org/abs/hep-ph/9309335} {arXiv:hep-ph/9309335} \BibitemShut
  {NoStop}%
\bibitem [{\citenamefont {Minkowski}(1977)}]{Minkowski:1977sc}%
  \BibitemOpen
  \bibfield  {author} {\bibinfo {author} {\bibfnamefont {P.}~\bibnamefont
  {Minkowski}},\ }\href {\doibase 10.1016/0370-2693(77)90435-X} {\bibfield
  {journal} {\bibinfo  {journal} {Phys. Lett. B}\ }\textbf {\bibinfo {volume}
  {67}},\ \bibinfo {pages} {421} (\bibinfo {year} {1977})}\BibitemShut
  {NoStop}%
\bibitem [{\citenamefont {Yanagida}(1979)}]{Yanagida:1979as}%
  \BibitemOpen
  \bibfield  {author} {\bibinfo {author} {\bibfnamefont {T.}~\bibnamefont
  {Yanagida}},\ }\href@noop {} {\bibfield  {journal} {\bibinfo  {journal}
  {Conf. Proc. C}\ }\textbf {\bibinfo {volume} {7902131}},\ \bibinfo {pages}
  {95} (\bibinfo {year} {1979})}\BibitemShut {NoStop}%
\bibitem [{\citenamefont {Mohapatra}\ and\ \citenamefont
  {Senjanovic}(1980)}]{Mohapatra:1979ia}%
  \BibitemOpen
  \bibfield  {author} {\bibinfo {author} {\bibfnamefont {R.~N.}\ \bibnamefont
  {Mohapatra}}\ and\ \bibinfo {author} {\bibfnamefont {G.}~\bibnamefont
  {Senjanovic}},\ }\href {\doibase 10.1103/PhysRevLett.44.912} {\bibfield
  {journal} {\bibinfo  {journal} {Phys. Rev. Lett.}\ }\textbf {\bibinfo
  {volume} {44}},\ \bibinfo {pages} {912} (\bibinfo {year} {1980})}\BibitemShut
  {NoStop}%
\bibitem [{\citenamefont {Schechter}\ and\ \citenamefont
  {Valle}(1980)}]{Schechter:1980gr}%
  \BibitemOpen
  \bibfield  {author} {\bibinfo {author} {\bibfnamefont {J.}~\bibnamefont
  {Schechter}}\ and\ \bibinfo {author} {\bibfnamefont {J.~W.~F.}\ \bibnamefont
  {Valle}},\ }\href {\doibase 10.1103/PhysRevD.22.2227} {\bibfield  {journal}
  {\bibinfo  {journal} {Phys. Rev. D}\ }\textbf {\bibinfo {volume} {22}},\
  \bibinfo {pages} {2227} (\bibinfo {year} {1980})}\BibitemShut {NoStop}%
\bibitem [{\citenamefont {Joshipura}\ and\ \citenamefont
  {Patel}(2011)}]{Joshipura:2011nn}%
  \BibitemOpen
  \bibfield  {author} {\bibinfo {author} {\bibfnamefont {A.~S.}\ \bibnamefont
  {Joshipura}}\ and\ \bibinfo {author} {\bibfnamefont {K.~M.}\ \bibnamefont
  {Patel}},\ }\href {\doibase 10.1103/PhysRevD.83.095002} {\bibfield  {journal}
  {\bibinfo  {journal} {Phys. Rev. D}\ }\textbf {\bibinfo {volume} {83}},\
  \bibinfo {pages} {095002} (\bibinfo {year} {2011})},\ \Eprint
  {http://arxiv.org/abs/1102.5148} {arXiv:1102.5148 [hep-ph]} \BibitemShut
  {NoStop}%
\bibitem [{\citenamefont {Mummidi}\ and\ \citenamefont
  {Patel}(2021)}]{Mummidi:2021anm}%
  \BibitemOpen
  \bibfield  {author} {\bibinfo {author} {\bibfnamefont {V.~S.}\ \bibnamefont
  {Mummidi}}\ and\ \bibinfo {author} {\bibfnamefont {K.~M.}\ \bibnamefont
  {Patel}},\ }\href {\doibase 10.1007/JHEP12(2021)042} {\bibfield  {journal}
  {\bibinfo  {journal} {JHEP}\ }\textbf {\bibinfo {volume} {12}},\ \bibinfo
  {pages} {042} (\bibinfo {year} {2021})},\ \Eprint
  {http://arxiv.org/abs/2109.04050} {arXiv:2109.04050 [hep-ph]} \BibitemShut
  {NoStop}%
\bibitem [{\citenamefont {Sartore}\ and\ \citenamefont
  {Schienbein}(2021)}]{Sartore:2020gou}%
  \BibitemOpen
  \bibfield  {author} {\bibinfo {author} {\bibfnamefont {L.}~\bibnamefont
  {Sartore}}\ and\ \bibinfo {author} {\bibfnamefont {I.}~\bibnamefont
  {Schienbein}},\ }\href {\doibase 10.1016/j.cpc.2020.107819} {\bibfield
  {journal} {\bibinfo  {journal} {Comput. Phys. Commun.}\ }\textbf {\bibinfo
  {volume} {261}},\ \bibinfo {pages} {107819} (\bibinfo {year} {2021})},\
  \Eprint {http://arxiv.org/abs/2007.12700} {arXiv:2007.12700 [hep-ph]}
  \BibitemShut {NoStop}%
\bibitem [{\citenamefont {Suryanarayana~Mummidi}\ and\ \citenamefont
  {Patel}(2020)}]{Surya:2020ydm}%
  \BibitemOpen
  \bibfield  {author} {\bibinfo {author} {\bibfnamefont {V.}~\bibnamefont
  {Suryanarayana~Mummidi}}\ and\ \bibinfo {author} {\bibfnamefont {K.~M.}\
  \bibnamefont {Patel}},\ }\href {\doibase 10.1103/PhysRevD.101.115008}
  {\bibfield  {journal} {\bibinfo  {journal} {Phys. Rev. D}\ }\textbf {\bibinfo
  {volume} {101}},\ \bibinfo {pages} {115008} (\bibinfo {year} {2020})},\
  \Eprint {http://arxiv.org/abs/2001.01505} {arXiv:2001.01505 [hep-ph]}
  \BibitemShut {NoStop}%
\bibitem [{\citenamefont {Allwicher}\ \emph {et~al.}(2021)\citenamefont
  {Allwicher}, \citenamefont {Arnan}, \citenamefont {Barducci},\ and\
  \citenamefont {Nardecchia}}]{Allwicher:2021rtd}%
  \BibitemOpen
  \bibfield  {author} {\bibinfo {author} {\bibfnamefont {L.}~\bibnamefont
  {Allwicher}}, \bibinfo {author} {\bibfnamefont {P.}~\bibnamefont {Arnan}},
  \bibinfo {author} {\bibfnamefont {D.}~\bibnamefont {Barducci}}, \ and\
  \bibinfo {author} {\bibfnamefont {M.}~\bibnamefont {Nardecchia}},\ }\href
  {\doibase 10.1007/JHEP10(2021)129} {\bibfield  {journal} {\bibinfo  {journal}
  {JHEP}\ }\textbf {\bibinfo {volume} {10}},\ \bibinfo {pages} {129} (\bibinfo
  {year} {2021})},\ \Eprint {http://arxiv.org/abs/2108.00013} {arXiv:2108.00013
  [hep-ph]} \BibitemShut {NoStop}%
\bibitem [{\citenamefont {Masiero}\ \emph {et~al.}(1982)\citenamefont
  {Masiero}, \citenamefont {Nanopoulos}, \citenamefont {Tamvakis},\ and\
  \citenamefont {Yanagida}}]{Masiero:1982fe}%
  \BibitemOpen
  \bibfield  {author} {\bibinfo {author} {\bibfnamefont {A.}~\bibnamefont
  {Masiero}}, \bibinfo {author} {\bibfnamefont {D.~V.}\ \bibnamefont
  {Nanopoulos}}, \bibinfo {author} {\bibfnamefont {K.}~\bibnamefont
  {Tamvakis}}, \ and\ \bibinfo {author} {\bibfnamefont {T.}~\bibnamefont
  {Yanagida}},\ }\href {\doibase 10.1016/0370-2693(82)90522-6} {\bibfield
  {journal} {\bibinfo  {journal} {Phys. Lett. B}\ }\textbf {\bibinfo {volume}
  {115}},\ \bibinfo {pages} {380} (\bibinfo {year} {1982})}\BibitemShut
  {NoStop}%
\bibitem [{\citenamefont {Grinstein}(1982)}]{Grinstein:1982um}%
  \BibitemOpen
  \bibfield  {author} {\bibinfo {author} {\bibfnamefont {B.}~\bibnamefont
  {Grinstein}},\ }\href {\doibase 10.1016/0550-3213(82)90275-9} {\bibfield
  {journal} {\bibinfo  {journal} {Nucl. Phys. B}\ }\textbf {\bibinfo {volume}
  {206}},\ \bibinfo {pages} {387} (\bibinfo {year} {1982})}\BibitemShut
  {NoStop}%
\bibitem [{\citenamefont {Babu}\ \emph {et~al.}(2007)\citenamefont {Babu},
  \citenamefont {Gogoladze},\ and\ \citenamefont
  {Tavartkiladze}}]{Babu:2006nf}%
  \BibitemOpen
  \bibfield  {author} {\bibinfo {author} {\bibfnamefont {K.~S.}\ \bibnamefont
  {Babu}}, \bibinfo {author} {\bibfnamefont {I.}~\bibnamefont {Gogoladze}}, \
  and\ \bibinfo {author} {\bibfnamefont {Z.}~\bibnamefont {Tavartkiladze}},\
  }\href {\doibase 10.1016/j.physletb.2007.02.050} {\bibfield  {journal}
  {\bibinfo  {journal} {Phys. Lett. B}\ }\textbf {\bibinfo {volume} {650}},\
  \bibinfo {pages} {49} (\bibinfo {year} {2007})},\ \Eprint
  {http://arxiv.org/abs/hep-ph/0612315} {arXiv:hep-ph/0612315} \BibitemShut
  {NoStop}%
\bibitem [{\citenamefont {Dorsner}(2012)}]{Dorsner:2012uz}%
  \BibitemOpen
  \bibfield  {author} {\bibinfo {author} {\bibfnamefont {I.}~\bibnamefont
  {Dorsner}},\ }\href {\doibase 10.1103/PhysRevD.86.055009} {\bibfield
  {journal} {\bibinfo  {journal} {Phys. Rev. D}\ }\textbf {\bibinfo {volume}
  {86}},\ \bibinfo {pages} {055009} (\bibinfo {year} {2012})},\ \Eprint
  {http://arxiv.org/abs/1206.5998} {arXiv:1206.5998 [hep-ph]} \BibitemShut
  {NoStop}%
\bibitem [{\citenamefont {Esteban}\ \emph {et~al.}(2020)\citenamefont
  {Esteban}, \citenamefont {Gonzalez-Garcia}, \citenamefont {Maltoni},
  \citenamefont {Schwetz},\ and\ \citenamefont {Zhou}}]{Esteban:2020cvm}%
  \BibitemOpen
  \bibfield  {author} {\bibinfo {author} {\bibfnamefont {I.}~\bibnamefont
  {Esteban}}, \bibinfo {author} {\bibfnamefont {M.~C.}\ \bibnamefont
  {Gonzalez-Garcia}}, \bibinfo {author} {\bibfnamefont {M.}~\bibnamefont
  {Maltoni}}, \bibinfo {author} {\bibfnamefont {T.}~\bibnamefont {Schwetz}}, \
  and\ \bibinfo {author} {\bibfnamefont {A.}~\bibnamefont {Zhou}},\ }\href
  {\doibase 10.1007/JHEP09(2020)178} {\bibfield  {journal} {\bibinfo  {journal}
  {JHEP}\ }\textbf {\bibinfo {volume} {09}},\ \bibinfo {pages} {178} (\bibinfo
  {year} {2020})},\ \Eprint {http://arxiv.org/abs/2007.14792} {arXiv:2007.14792
  [hep-ph]} \BibitemShut {NoStop}%
\bibitem [{\citenamefont {Chankowski}\ and\ \citenamefont
  {Pluciennik}(1993)}]{Chankowski:1993tx}%
  \BibitemOpen
  \bibfield  {author} {\bibinfo {author} {\bibfnamefont {P.~H.}\ \bibnamefont
  {Chankowski}}\ and\ \bibinfo {author} {\bibfnamefont {Z.}~\bibnamefont
  {Pluciennik}},\ }\href {\doibase 10.1016/0370-2693(93)90330-K} {\bibfield
  {journal} {\bibinfo  {journal} {Phys. Lett. B}\ }\textbf {\bibinfo {volume}
  {316}},\ \bibinfo {pages} {312} (\bibinfo {year} {1993})},\ \Eprint
  {http://arxiv.org/abs/hep-ph/9306333} {arXiv:hep-ph/9306333} \BibitemShut
  {NoStop}%
\bibitem [{\citenamefont {Babu}\ \emph {et~al.}(1993)\citenamefont {Babu},
  \citenamefont {Leung},\ and\ \citenamefont {Pantaleone}}]{Babu:1993qv}%
  \BibitemOpen
  \bibfield  {author} {\bibinfo {author} {\bibfnamefont {K.~S.}\ \bibnamefont
  {Babu}}, \bibinfo {author} {\bibfnamefont {C.~N.}\ \bibnamefont {Leung}}, \
  and\ \bibinfo {author} {\bibfnamefont {J.~T.}\ \bibnamefont {Pantaleone}},\
  }\href {\doibase 10.1016/0370-2693(93)90801-N} {\bibfield  {journal}
  {\bibinfo  {journal} {Phys. Lett. B}\ }\textbf {\bibinfo {volume} {319}},\
  \bibinfo {pages} {191} (\bibinfo {year} {1993})},\ \Eprint
  {http://arxiv.org/abs/hep-ph/9309223} {arXiv:hep-ph/9309223} \BibitemShut
  {NoStop}%
\bibitem [{\citenamefont {Antusch}\ \emph {et~al.}(2005)\citenamefont
  {Antusch}, \citenamefont {Kersten}, \citenamefont {Lindner}, \citenamefont
  {Ratz},\ and\ \citenamefont {Schmidt}}]{Antusch:2005gp}%
  \BibitemOpen
  \bibfield  {author} {\bibinfo {author} {\bibfnamefont {S.}~\bibnamefont
  {Antusch}}, \bibinfo {author} {\bibfnamefont {J.}~\bibnamefont {Kersten}},
  \bibinfo {author} {\bibfnamefont {M.}~\bibnamefont {Lindner}}, \bibinfo
  {author} {\bibfnamefont {M.}~\bibnamefont {Ratz}}, \ and\ \bibinfo {author}
  {\bibfnamefont {M.~A.}\ \bibnamefont {Schmidt}},\ }\href {\doibase
  10.1088/1126-6708/2005/03/024} {\bibfield  {journal} {\bibinfo  {journal}
  {JHEP}\ }\textbf {\bibinfo {volume} {03}},\ \bibinfo {pages} {024} (\bibinfo
  {year} {2005})},\ \Eprint {http://arxiv.org/abs/hep-ph/0501272}
  {arXiv:hep-ph/0501272} \BibitemShut {NoStop}%
\bibitem [{\citenamefont {Mei}(2005)}]{Mei:2005qp}%
  \BibitemOpen
  \bibfield  {author} {\bibinfo {author} {\bibfnamefont {J.-w.}\ \bibnamefont
  {Mei}},\ }\href {\doibase 10.1103/PhysRevD.71.073012} {\bibfield  {journal}
  {\bibinfo  {journal} {Phys. Rev. D}\ }\textbf {\bibinfo {volume} {71}},\
  \bibinfo {pages} {073012} (\bibinfo {year} {2005})},\ \Eprint
  {http://arxiv.org/abs/hep-ph/0502015} {arXiv:hep-ph/0502015} \BibitemShut
  {NoStop}%
\bibitem [{\citenamefont {Diaz-Cruz}\ \emph {et~al.}(2002)\citenamefont
  {Diaz-Cruz}, \citenamefont {Murayama},\ and\ \citenamefont
  {Pierce}}]{Diaz-Cruz:2000nvf}%
  \BibitemOpen
  \bibfield  {author} {\bibinfo {author} {\bibfnamefont {J.~L.}\ \bibnamefont
  {Diaz-Cruz}}, \bibinfo {author} {\bibfnamefont {H.}~\bibnamefont {Murayama}},
  \ and\ \bibinfo {author} {\bibfnamefont {A.}~\bibnamefont {Pierce}},\ }\href
  {\doibase 10.1103/PhysRevD.65.075011} {\bibfield  {journal} {\bibinfo
  {journal} {Phys. Rev. D}\ }\textbf {\bibinfo {volume} {65}},\ \bibinfo
  {pages} {075011} (\bibinfo {year} {2002})},\ \Eprint
  {http://arxiv.org/abs/hep-ph/0012275} {arXiv:hep-ph/0012275} \BibitemShut
  {NoStop}%
\bibitem [{\citenamefont {Enkhbat}(2009)}]{Enkhbat:2009jt}%
  \BibitemOpen
  \bibfield  {author} {\bibinfo {author} {\bibfnamefont {T.}~\bibnamefont
  {Enkhbat}},\ }\href@noop {} {\  (\bibinfo {year} {2009})},\ \Eprint
  {http://arxiv.org/abs/0909.5597} {arXiv:0909.5597 [hep-ph]} \BibitemShut
  {NoStop}%
\bibitem [{\citenamefont {Oliensis}\ and\ \citenamefont
  {Fischler}(1983)}]{PhysRevD.28.194}%
  \BibitemOpen
  \bibfield  {author} {\bibinfo {author} {\bibfnamefont {J.}~\bibnamefont
  {Oliensis}}\ and\ \bibinfo {author} {\bibfnamefont {M.}~\bibnamefont
  {Fischler}},\ }\href {\doibase 10.1103/PhysRevD.28.194} {\bibfield  {journal}
  {\bibinfo  {journal} {Phys. Rev. D}\ }\textbf {\bibinfo {volume} {28}},\
  \bibinfo {pages} {194} (\bibinfo {year} {1983})}\BibitemShut {NoStop}%
\bibitem [{\citenamefont {Kane}\ \emph {et~al.}(1994)\citenamefont {Kane},
  \citenamefont {Kolda}, \citenamefont {Roszkowski},\ and\ \citenamefont
  {Wells}}]{Kane:1993td}%
  \BibitemOpen
  \bibfield  {author} {\bibinfo {author} {\bibfnamefont {G.~L.}\ \bibnamefont
  {Kane}}, \bibinfo {author} {\bibfnamefont {C.~F.}\ \bibnamefont {Kolda}},
  \bibinfo {author} {\bibfnamefont {L.}~\bibnamefont {Roszkowski}}, \ and\
  \bibinfo {author} {\bibfnamefont {J.~D.}\ \bibnamefont {Wells}},\ }\href
  {\doibase 10.1103/PhysRevD.49.6173} {\bibfield  {journal} {\bibinfo
  {journal} {Phys. Rev. D}\ }\textbf {\bibinfo {volume} {49}},\ \bibinfo
  {pages} {6173} (\bibinfo {year} {1994})},\ \Eprint
  {http://arxiv.org/abs/hep-ph/9312272} {arXiv:hep-ph/9312272} \BibitemShut
  {NoStop}%
\bibitem [{\citenamefont {Weinberg}(1980)}]{Weinberg:1980wa}%
  \BibitemOpen
  \bibfield  {author} {\bibinfo {author} {\bibfnamefont {S.}~\bibnamefont
  {Weinberg}},\ }\href {\doibase 10.1016/0370-2693(80)90660-7} {\bibfield
  {journal} {\bibinfo  {journal} {Phys. Lett. B}\ }\textbf {\bibinfo {volume}
  {91}},\ \bibinfo {pages} {51} (\bibinfo {year} {1980})}\BibitemShut {NoStop}%
\bibitem [{\citenamefont {Hall}(1981)}]{Hall:1980kf}%
  \BibitemOpen
  \bibfield  {author} {\bibinfo {author} {\bibfnamefont {L.~J.}\ \bibnamefont
  {Hall}},\ }\href {\doibase 10.1016/0550-3213(81)90498-3} {\bibfield
  {journal} {\bibinfo  {journal} {Nucl. Phys. B}\ }\textbf {\bibinfo {volume}
  {178}},\ \bibinfo {pages} {75} (\bibinfo {year} {1981})}\BibitemShut
  {NoStop}%
\end{thebibliography}%
\bibliographystyle{apsrev4-1}

\pagebreak
\onecolumngrid

\appendix
\section{General formula of 1-loop matching}
\label{app:gen_form}
In this Appendix, we show an explicit derivation of the 1-loop matching relation, Eq. (\ref{dY_gen}), for the Yukawa coupling matrices. Computation of Yukawa thresholds is carried out for the first time in \cite{PhysRevD.28.194,Kane:1993td,Hempfling:1993kv,Wright:1994qb} following the analogous procedure developed for the gauge couplings in \cite{Weinberg:1980wa,Hall:1980kf}. We closely follow \cite{Wright:1994qb} and outline the treatment for the Yukawa couplings for completeness.  

Consider chiral fermions $\psi_i$, $\chi_i$ and a scalar $\phi$ with the following gauge and Yukawa interactions in the full theory,
\be \label{full_l}
{\cal L} = i \overline{\psi}_i \slashed{D} \psi_i + i \overline{\chi}_i \slashed{D} \chi_i + D_\mu \phi^\dagger D^\mu \phi - \left\{Y_{ij} \psi_i^T C \chi_j \phi + {\rm h.c.} \right\}\,.\ee
Let $\psi_i$, $\chi_i$ and $\phi$ decompose in the light fields, namely $\psi_{li}$, $\chi_{li}$ and $\phi_l$, and the heavy fields $\psi_{h i}$, $\chi_{h i}$ and $\phi_h$, respectively. Integrating out the heavy fields, the effective Lagrangian of the light fields is given by
\be \label{light_L}
{\cal L}_{\rm eff} = i \overline{\psi}_{l i} \slashed{D} (Z_\psi)_{ij} \psi_{l j} + i \overline{\chi}_{li} \slashed{D} (Z_\psi)_{ij} \chi_{lj} + D_\mu \phi_l^\dagger Z_\phi D^\mu \phi_l - \left\{\tilde{Y}_{ij} \psi_{li}^T C \chi_{lj} \phi_l + {\rm h.c.} \right\} + ...\,,\ee
where ... denotes the non-renormalizable operators induced because of the integrated out fields. The $Z$-parameters can be parametrized as
\be \label{Z}
(Z_{\psi,\chi})_{ij} = \delta_{ij} + (K_{\psi,\chi})_{ij}\,,~~Z_\phi = 1 + K_\phi\,,\ee
where $K_{\psi,\chi,\phi}$ can be evaluated using the wavefunction renormalization of the corresponding light field at 1-loop involving at least one heavy field in the loop. Similarly, $\tilde{Y}$ in Eq. (\ref{light_L}) can be written as
\be \label{Y_tilde}
\tilde{Y} = Y + \delta Y\,,\ee
where $\delta Y$ is the 1-loop Yukawa vertex correction with heavy fields in the loop.

Canonical normalization of the kinetic terms requires field redefinitions. To achieve this, we define
\beqa \label{filed_red}
\psi_{li} = \left(U_\psi \tilde{Z}_\psi^{-1/2} U_\psi^\dagger \right)_{ij}\, \tilde{\psi}_{l j}\,,~~\chi_{li} = \left(U_\chi \tilde{Z}_\chi^{-1/2} U_\chi^\dagger \right)_{ij}\, \tilde{\chi}_{l j}\,,~~\phi_l = Z_\phi^{-1/2}\,\tilde{\phi}_l\,.\eeqa
Here, $\tilde{Z}_{\psi,\chi} = U_{\psi,\chi}^\dagger Z_{\psi,\chi} U_{\psi,\chi}$ are diagonal matrices. Substitution of the above in Eq. (\ref{light_L}) leads to canonically normalised kinetic terms for the light fermions $\tilde{\psi}_{li}$, $\tilde{\chi}_{li}$ and the scalar $\tilde{\phi}_l$ as it can be verified easily. Further, the effective Yukawa couplings in the new basis can be determined by substituting Eq. (\ref{filed_red}) in the last term of Eq. (\ref{light_L}). We then find
\be \label{eff_LY}
{\cal L}_{\rm eff} \supset (Y_{\rm eff})_{ij}\, \tilde{\psi}_{li}^T C \tilde{\chi}_{lj} \phi_l + {\rm h.c.}\,, \ee
with
\beqa \label{Y_eff_1}
Y_{\rm eff} &=& U_\psi^* \tilde{Z}_\psi^{-1/2} U_\psi^T\,  \tilde{Y}\, U_\chi \tilde{Z}_\chi^{-1/2} U_\chi^\dagger\,Z_\phi^{-1/2}\,. \eeqa
Using the definitions, Eq. (\ref{Z}), one can express
\be \label{}
\tilde{Z}_{\psi,\chi}^{-1/2} = ({\bf 1} + \tilde{K}_{\psi,\chi})^{-1/2} \simeq {\bf 1} - \frac{1}{2} \tilde{K}_{\psi,\chi}\,, \ee
where $\tilde{K}_{\psi,\chi} = U_{\psi,\chi}^\dagger K_{\psi,\chi} U_{\psi,\chi}$ are diagonal and real matrices with $(\tilde{K}_{\psi,\chi})_{ii} < 1$. Similarly, $Z_\phi^{-1/2} \simeq 1-\frac{1}{2} K_\phi$. Substituting these in Eq. (\ref{Y_eff_1}) and keeping only the leading order terms in $\delta Y$ and $K$, we find
\beqa \label{Y_eff_2}
Y_{\rm eff} & =  & Y \left(1-\frac{1}{2}K_\phi \right) + \delta Y -\frac{1}{2} K_\psi^T Y - \frac{1}{2} Y K_\chi\,. \eeqa
The above can be used as a 1-loop corrected matching condition at a renormalization scale $\mu$ by replacing $\delta Y$ and $K_{\psi,\chi,\phi}$ by their finite parts defined in the $\overline{\rm MS}$ scheme. Eq. (\ref{Y_eff_2}) then provides the 1-loop corrected expression for the effective Yukawa couplings in terms of the original Yukawa couplings of the full theory and the leading corrections arising from the heavy particles.

\section{Loop integration factors}
\label{app:LF}
The loop integration factors appearing in Eqs. (\ref{dY}) and (\ref{K}) are given by
\beqa \label{LF_f}
f[m_1^2,m_2^2] &=& -\frac{1}{16 \pi^2} \left(\frac{m_1^2 \log \frac{m_1^2}{\mu^2}-m_2^2\log \frac{m_2^2}{\mu^2}}{m_1^2-m_2^2} - 1 \right)\,, \eeqa
\beqa \label{LF_h}
h[m_1^2,m_2^2] &=& \frac{1}{16 \pi^2} \left(\frac{1}{2} \log\frac{m_1^2}{\mu^2} +  \frac{\frac{1}{2} r^2 \log r -\frac{3}{4}r^2 + r - \frac{1}{4}}{(1-r)^2}\right)\,, \eeqa
and
\beqa \label{LF_g}
g[m_1^2,m_2^2] & = & \frac{1}{16 \pi^2} \frac{\frac{r^3}{6}-r^2+\frac{r}{2}+ r \log r + \frac{1}{3}}{(1 -r)^3}\,,\eeqa
where $r=m_2^2/m_1^2$ in the last two equations.

\section{Example numerical solutions}
\label{app:sol}

In this appendix, we give the fitted values of the Yukawa coupling matrices $Y_{1,2,3}$ corresponding to the benchmark solutions I, II and III, as listed in Table \ref{tab:tab1}, obtained at $\mu=10^{16}$ GeV. For Solution I, we find 
\beqa
Y_1 &=&
\left(
\begin{array}{ccc}
 2.79 \times 10^{-6} & 0 & 0 \\
 0 & 1.41 \times 10^{-3} & 0 \\
 0 & 0 & 4.36\times 10 ^{-1}  \\
\end{array}
\right)\,,~~~
Y_3 = \left(
\begin{array}{c}
 6.63 \times 10^{-2}  \\
 -3.46  \\
 2.93  \\
\end{array}
\right)\,, \nonumber\\
Y_2 &=& \left(
\begin{array}{ccc}
 -4.24\times 10^{-6}-i\,1.07\times 10^{-5}  & -1.84\times 10^{-4}+i\,4.81\times 10^{-5}  & 1.24 \times 10^{-4} -i\,4.0\times 10^{-5}  \\
 8.98\times 10^{-5} + i\, 7.71\times 10^{-5} & 2.76\times10^{-4} -i\,7.98\times10^{-5} & -3.57\times 10^{-4}+i\,3.84\times 10^{-4} \\
 -7.58\times10^{-4}-i\,7.53\times 10^{-4}  & 8.60\times10^{-3} -i\,1.45\times 10^{-3} & -1.15\times10^{-3}-i\,2.55\times10^{-3} \\
\end{array}
\right)\,.\eeqa
In the case of Solution II, we get
\beqa
Y_1&=&\left(
\begin{array}{ccc}
 2.78 \times 10^{-6} & 0 & 0 \\
 0 & 1.40\times 10^{-3} & 0 \\
 0 & 0 & -4.24\times 10^{-1} \\
\end{array}
\right)\,,~~~  Y_3 = \left(
\begin{array}{cc}
 2.96\times 10^{-1} & -2.43\times 10^{-2} \\
 -3.44 & 8.92\times 10^{-3} \\
 -3.50 & -4.25\times 10^{-2} \\
\end{array}
\right)\,, \nonumber\\
Y_2 &=& \left(
\begin{array}{ccc}
 -1.54\times 10^{-6}+i\,4.92\times 10^{-6}  & -2.65\times 10^{-5}-i\,1.05\times 10^{-4}  & 1.39 \times 10^{-5} -i\,1.20\times 10^{-4}  \\
 -2.45\times 10^{-5} - i\, 3.34\times 10^{-5} & 9.05\times10^{-4} +i\,2.73\times10^{-5} & 5.95\times 10^{-4}-i\,1.33\times 10^{-4} \\
 3.31\times10^{-4}+i\,4.24\times 10^{-4}  & -6.19\times10^{-3} +i\,4.74\times 10^{-3} & 2.15\times10^{-3}+i\,3.98\times10^{-3} \\
\end{array}
\right)\,.\eeqa
Similarly, for Solution III, we find
\beqa
Y_1&=&\left(
\begin{array}{ccc}
 2.79 \times 10^{-6} & 0 & 0 \\
 0 & 1.41 \times 10^{-3} & 0 \\
 0 & 0 & -4.28\times 10^{-1} \\
\end{array}
\right)\,,~~~  Y_3 = \left(
\begin{array}{ccc}
 3.40 \times 10^{-2}  & 9.83 \times 10^{-2} & 1.98 \times 10^{-2} \\
 -4.23\times 10^{-3} & 3.50 & -2.64 \times 10^{-1} \\
 -5.33\times 10^{-2} & -3.44 &9.50\times 10^{-2} \\
\end{array}
\right)\,, \nonumber\\
Y_2 &=& \left(
\begin{array}{ccc}
 6.10\times 10^{-6}-i\,6.22\times 10^{-7}  & 4.33\times 10^{-5}-i\,1.62\times 10^{-4}  & -3.93 \times 10^{-5} +i\,1.27\times 10^{-4}  \\
 1.26\times 10^{-5} + i\, 4.40\times 10^{-5} & 8.54\times10^{-4} -i\,4.19\times10^{-4} & -5.30\times 10^{-4}+i\,5.20\times 10^{-4} \\
 2.26\times10^{-4}-i\,4.77\times 10^{-4}  & -7.03\times10^{-3} +i\,9.19\times 10^{-4} & 8.49\times10^{-4}-i\,5.74\times10^{-3} \\
\end{array}
\right)\,.\eeqa

For all the solutions, we have used $g=0.53$ as the value of unified coupling at the GUT scale and $M_X = 10^{16}$ GeV. The optimized values of the masses of the colour triplet scalar and gauge singlet fermions are listed in Table \ref{tab:tab1} in the case of each solution.

\end{document}